\documentclass[journal]{IEEEtran}

\usepackage{framed,multirow}

\usepackage{amssymb}
\usepackage{latexsym}
\usepackage{epsfig}
\usepackage{graphicx}
\usepackage{amsmath}

\usepackage{url}
\usepackage{xcolor}

\usepackage{hyperref}

\begin{document}

% \verso{Dwarikanath Mahapatra \textit{et~al.}}

% \begin{frontmatter}

%%%%%%%%% TITLE
\title{CT Image Synthesis Using Weakly Supervised Segmentation and Geometric Inter-Label Relations For COVID Image Analysis}

% \author{Dwarikanath Mahapatra \\
% \and
% Inception Institute of Artificial Intelligence, Abu Dhabi,UAE
% }

\author{Dwarikanath Mahapatra$^{1}$,
\and
Ankur Singh$^{2}$ \\
\and
$^{1}$ Inception Institute of Artificial Intelligence, Abu Dhabi,  UAE \\
$^{2}$ Indian Institute of Technology, Kanpur, India \\ 
}

% \author{Dwarikanath Mahapatra$^{1}$,
% \and
% Ankur Singh$^{2}$,
% \and
% Behzad Bozorgtabar$^{3,4,5}$ \\
% \and
% $^{1}$ IIAI, Abu Dhabi \quad \quad 
% $^{2}$ Indian Institute of Technology, Kanpur, India \quad \quad
% $^{3}$ LTS5, EPFL, Lausanne \quad \quad \\
% $^{4}$ CIBM, Laussane \quad \quad
% $^{5}$ CHUV, Lausanne \\
% }

% \author[1]{Dwarikanath \snm{Mahapatra}\corref{cor1}}
% \ead{dwarikanath.mahapatra@inceptioniai.org}
% \cortext[cor1]{Corresponding author: 
%   Tel.: +971-58-546-1774}%;  
% %   fax: +0-000-000-0000;}
% % \author[2]{Ankur  \snm{Singh}} %\fnref{fn1}}
% % %
% % \author[3]{Behzad  \snm{Bozorgtabar}} %\fnref{fn1}}
% %
% % \fntext[fn1]{This is author footnote for second author.}
% % \author[1]{Ling  \snm{Shao}}
% %% Third author's email
% % \ead{author3@author.com}

% \address[1]{Inception Institute of AI, Abu Dhabi, UAE}
% % \address[2]{Indian Institute of Technology, Kanpur, India}
% % \address[3]{\'Ecole Polytechnique F\'ed\'erale de Lausanne (EPFL), Laussane, Switzerland}

% \communicated{S. Sarkar}

\maketitle

%%%%%%%%% ABSTRACT
\begin{abstract}
  While medical image segmentation is an important task for computer aided diagnosis, the high expertise requirement for pixelwise manual annotations makes it a challenging and time consuming task. Since conventional data augmentations do not fully represent the underlying distribution of the training set, the trained models have varying performance when tested on images captured from different sources. Most prior work on image synthesis for data augmentation ignore the interleaved geometric relationship between different anatomical labels. We propose improvements over previous GAN-based medical image synthesis methods by learning the relationship between different anatomical labels. We use a weakly supervised segmentation method to obtain pixel level semantic label map of images which is used learn the intrinsic relationship of geometry and shape across semantic labels. Latent space variable sampling results in diverse generated images from a base image and improves robustness. We use the synthetic images from our method to train networks for segmenting COVID-19 infected areas from lung CT images. The proposed method outperforms state-of-the-art segmentation methods on a public dataset. Ablation studies also demonstrate benefits of integrating geometry and diversity.
\end{abstract}

% \begin{keyword}
% % MSC codes here, in the form: \MSC code \sep code
% % or \MSC[2008] code \sep code (2000 is the default)
% % \MSC 41A05\sep 41A10\sep 65D05\sep 65D17
% %% Keywords
% \KWD Geometry \sep Augmentation \sep COVID \sep Classification
% \end{keyword}

% \end{frontmatter}

%%%%%%%%% BODY TEXT

\section{Introduction}
\label{sec:intro}

The novel Coronavirus Disease (COVID-19) pandemic has had a significant adverse impact on the global stage since the first reported cases in December 2019 \cite{InfNet_1,InfNet_2}. It has infected more than $4.5$ million people resulting in more than $315,000$ deaths across $210$ countries. %  As per the statistics released by the Center for Systems Science and Engineering (CSSE) at Johns Hopkins University (JHU) \cite{InfNet_3} (updated ----- ) the global total for COVID-19 cases is *******, which includes  ********* deaths in more than **** countries/regions. 
The gold standard for COVID-19 screening is the reverse-transcription polymerase chain reaction (RT-PCR) test. Equipment shortage
 and strict testing requirements 
limit rapid and accurate screening of the general populace, in addition to reports of RT-PCR testing exhibiting  high
false negative rates \cite{InfNet_4,Frontiers2020,Mahapatra_PR2020,ZGe_MTA2019,Behzad_PR2020,Mahapatra_CVIU2019,Mahapatra_CMIG2019,KuanarVC,MahapatraTMI2021,JuJbhi2020}. Radiological imaging such as  Xrays
and computed tomography (CT) have emerged as a useful tool in early COVID-19 screening  by
achieving high sensitivity (with RT-PCR results as reference) \cite{InfNet_4,Mahapatra_LME_PR2017,Zilly_CMIG_2016,Mahapatra_SSLAL_CD_CMPB,Mahapatra_SSLAL_Pro_JMI,Mahapatra_LME_CVIU,LiTMI_2015} and demonstrating robustness in diagnosis, follow-up
assessment, and evaluation of disease evolution \cite{InfNet_5,MahapatraJDI_Cardiac_FSL,Mahapatra_JSTSP2014,MahapatraTIP_RF2014,MahapatraTBME_Pro2014,MahapatraTMI_CD2013,MahapatraJDICD2013}. 

\begin{figure*}[t]
 \centering
\begin{tabular}{ccccc}
\includegraphics[height=3cm, width=3cm]{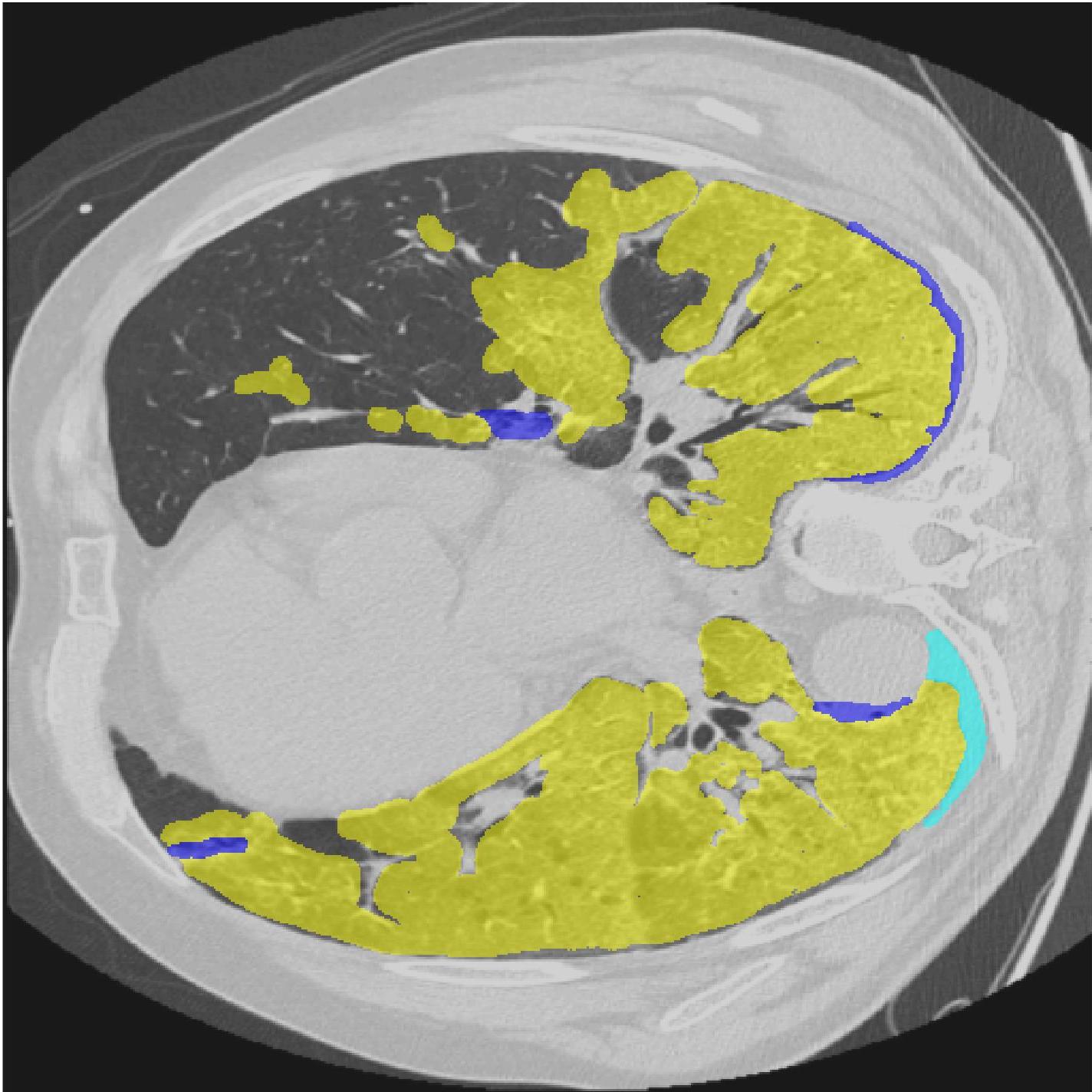} &
\includegraphics[height=3cm, width=3cm]{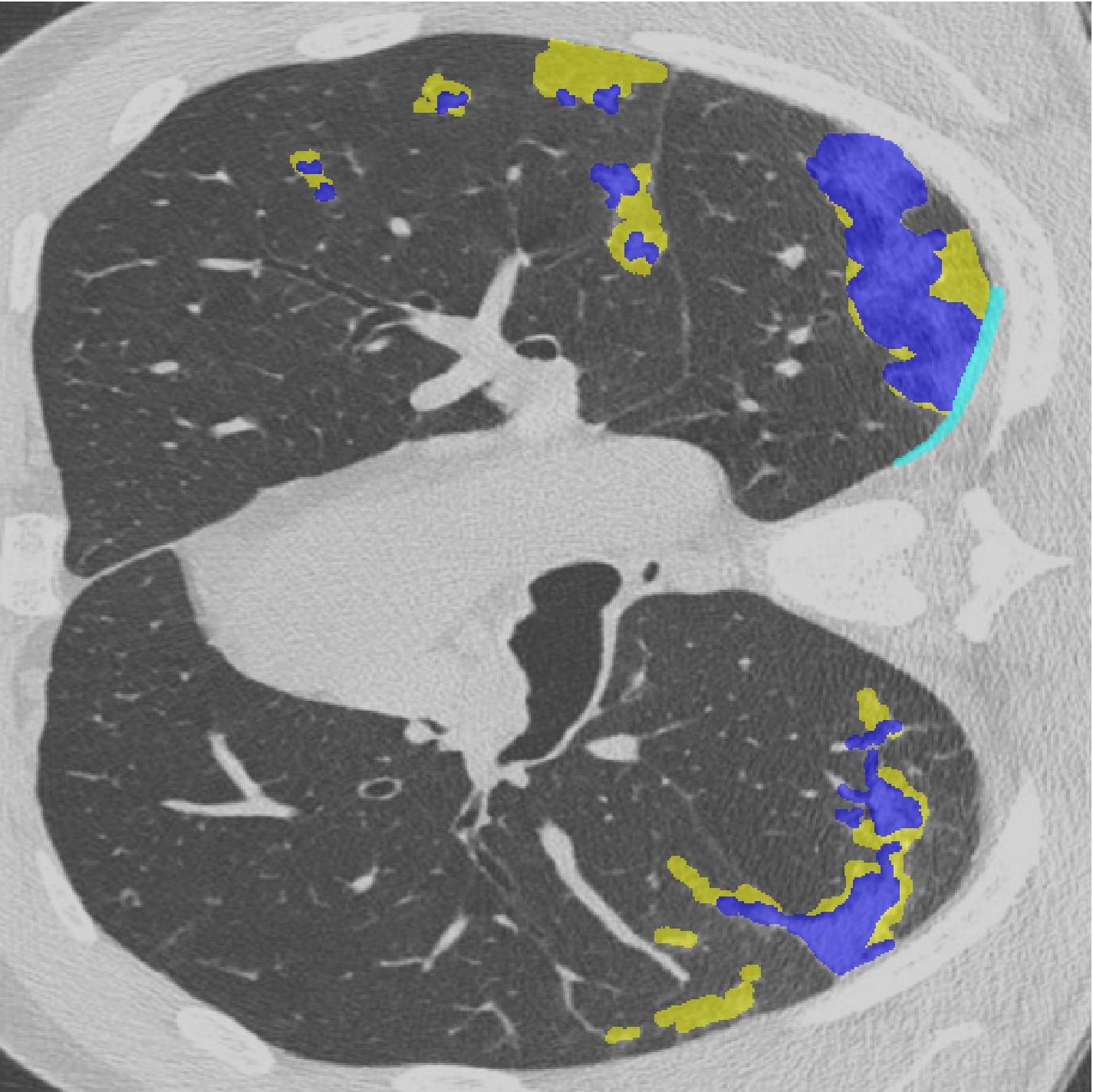} &
\includegraphics[height=3cm, width=3cm]{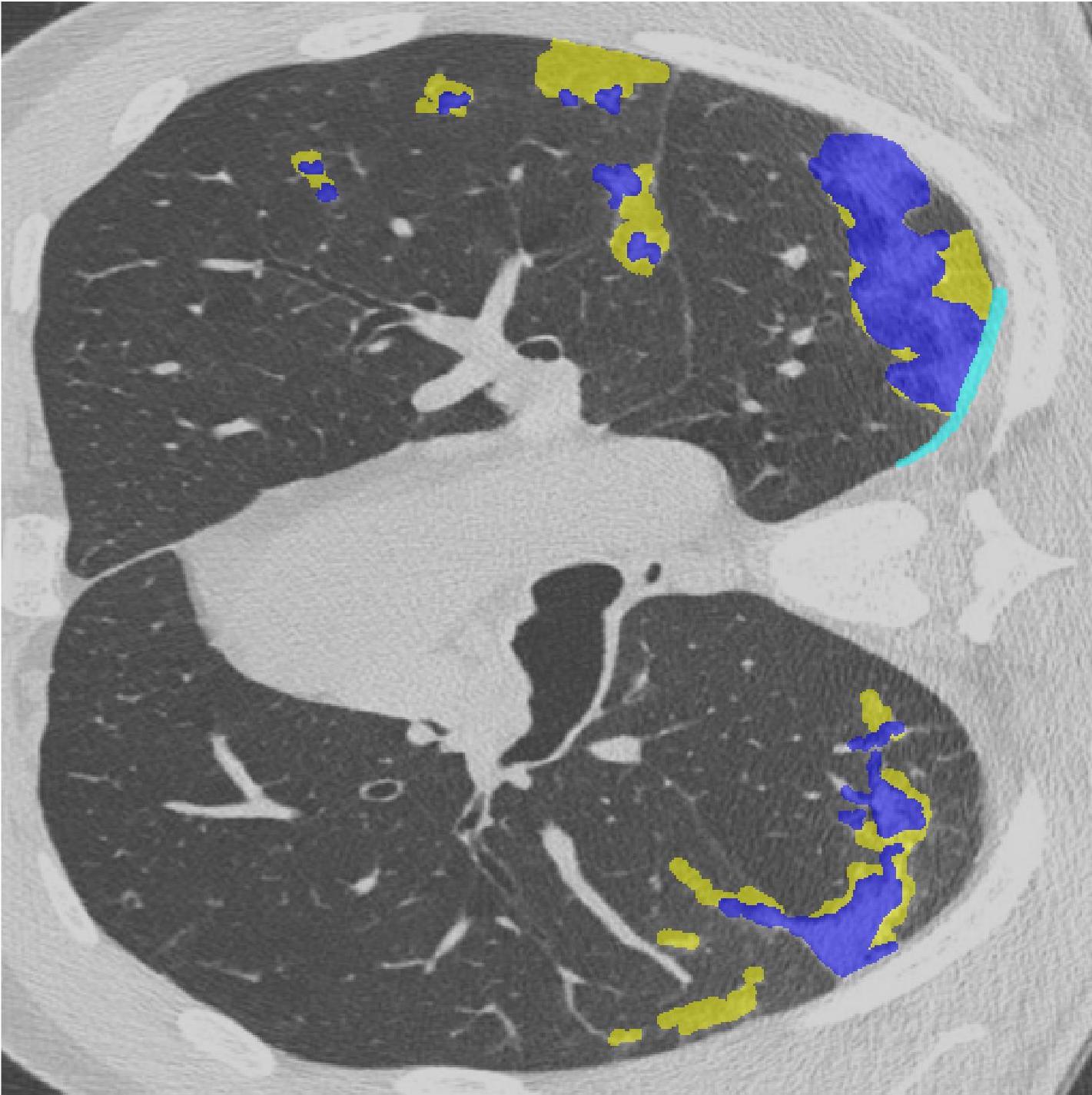} &
\includegraphics[height=3cm, width=3cm]{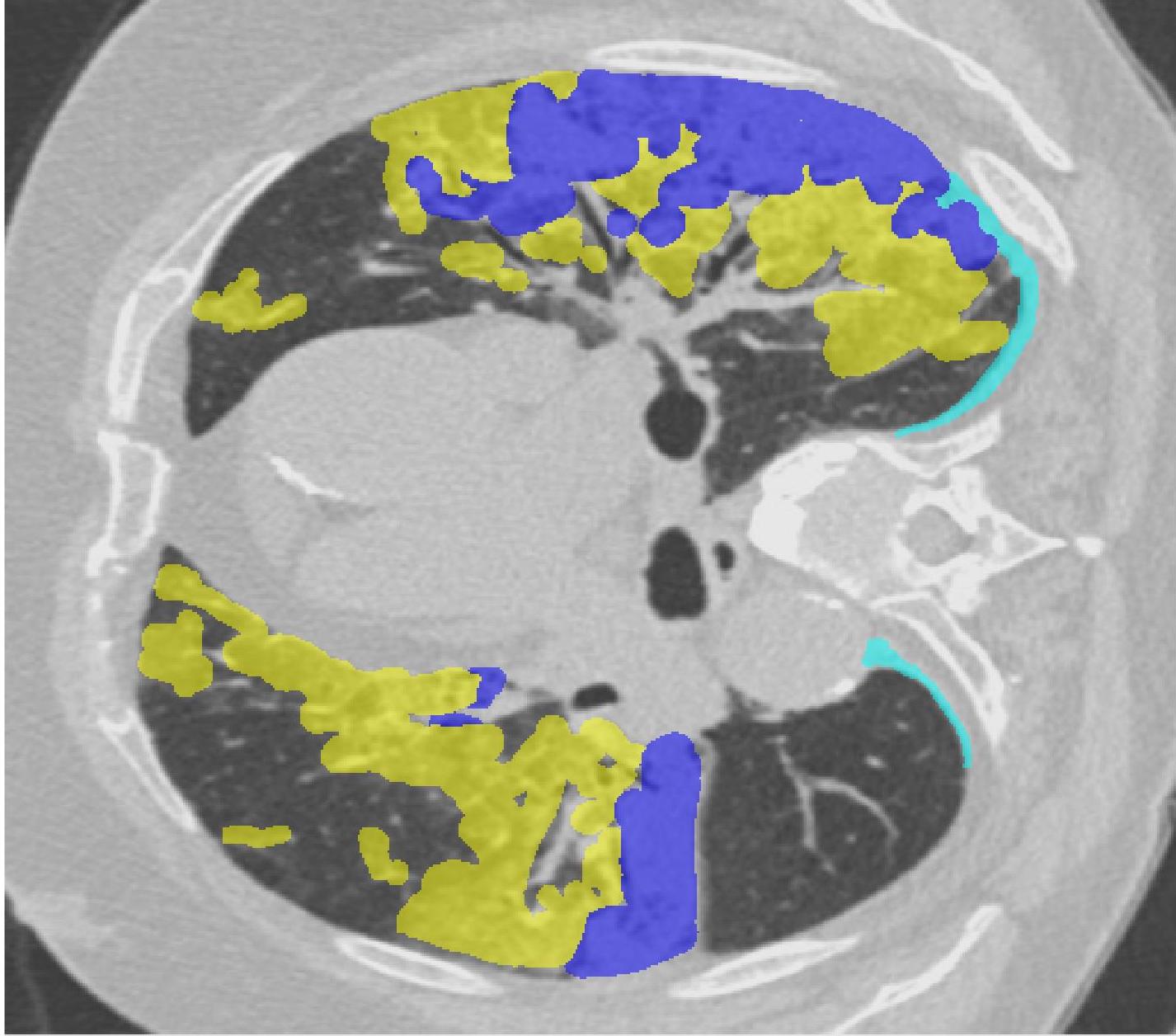} &
\includegraphics[height=3cm, width=3cm]{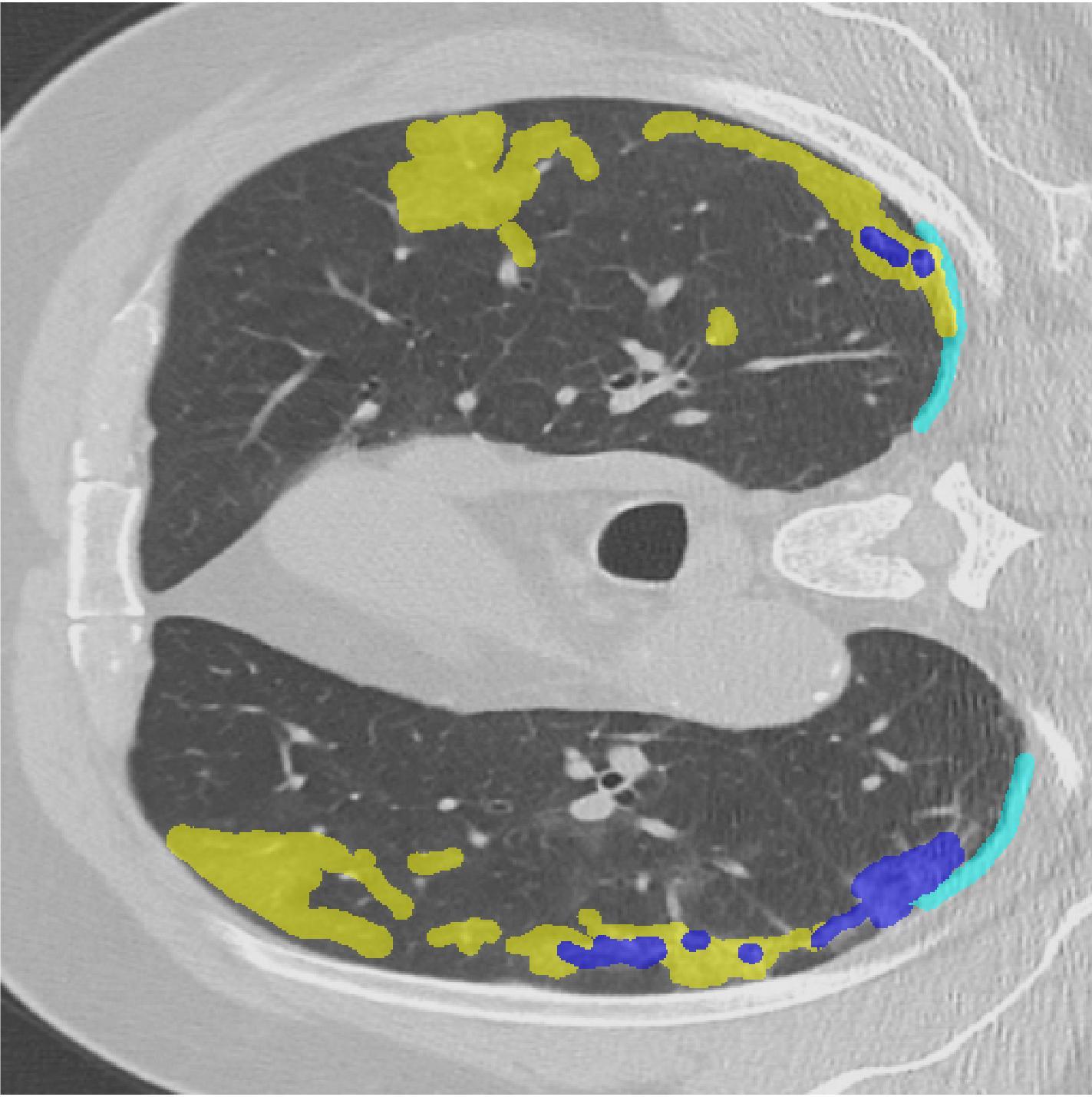} \\
(a) & (b)& (c) & (d) & (e) \\
\end{tabular}
\caption{Example of images showing disease pathologies such as ground glass opacity, consolidation and pleural effusion. }
\label{fig:disImage}
\end{figure*}

Although X-rays can be quickly acquired, CT screening provides a richer 3D view of the lung better suited for diagnosis. 
Recent studies of \cite{InfNet_4,InfNet_10,MahapatraJDIMutCont2013,MahapatraJDIGCSP2013,MahapatraJDIJSGR2013,MahapatraJDISkull2012,MahapatraTIP2012,MahapatraTBME2011,MahapatraEURASIP2010} provide evidence that CT scans (Figure~\ref{fig:disImage}) can be used to identify COVID-19 biomarkers such as  ground-glass opacity (GGO) in the early stage, and pulmonary consolidation in the late stage.
Thus qualitative evaluation of infection and longitudinal changes in CT scans can provide useful and important information for detecting COVID-19.

Recent methods such as \cite{InfNet_6,InfNet_15,Mahapatra_CVPR2020,Kuanar_ICIP19,Bozorgtabar_ICCV19,Xing_MICCAI19,Mahapatra_ISBI19,MahapatraAL_MICCAI18,Mahapatra_MLMI18,Sedai_OMIA18,TongDART20,Mahapatra_MICCAI20,Behzad_MICCAI20} have proposed deep learning (DL) systems
to detect COVID-19 patients from CT/Xray.
Wang et al. proposed COVID-Net  to identify
COVID-19 cases from chest xrays \cite{InfNet_16,Sedai_MLMI18,MahapatraGAN_ISBI18,Sedai_MICCAI17,Mahapatra_MICCAI17,Roy_ISBI17,Roy_DICTA16,Tennakoon_OMIA16,Sedai_OMIA16}. \cite{InfNet_17,Mahapatra_MLMI16,Sedai_EMBC16,Mahapatra_EMBC16,Mahapatra_MLMI15_Optic,Mahapatra_MLMI15_Prostate,Mahapatra_OMIA15,MahapatraISBI15_Optic,MahapatraISBI15_JSGR} proposed an
anomaly detection model to assist radiologists
in analyzing a large databse of chest X-ray images. \cite{InfNet_18,MahapatraISBI15_CD,KuangAMM14,Mahapatra_ABD2014,Schuffler_ABD2014,MahapatraISBI_CD2014,MahapatraMICCAI_CD2013,Schuffler_ABD2013,MahapatraProISBI13} developed a location-attention oriented model to calculate the infection probability of COVID-19 from CT images while a weakly-supervised DL method was developed in \cite{InfNet_19,MahapatraRVISBI13,MahapatraWssISBI13,MahapatraCDFssISBI13,MahapatraCDSPIE13,MahapatraABD12,MahapatraMLMI12,MahapatraSTACOM12,VosEMBC} for 3D CT volumes.

In comparison to classification (or diagnosis) related works, 
segmentation of pathological regions (e.g., infection areas) has received less attention [\cite{InfNet_20,InfNet_21,MahapatraGRSPIE12,MahapatraMiccaiIAHBD11,MahapatraMiccai11,MahapatraMiccai10,MahapatraICIP10,MahapatraICDIP10a,MahapatraICDIP10b,MahapatraMiccai08}]. 
Segmentation of infected regions from CT is challenging due to: 1) high variation in texture, size and position
of infections in CT scans. For
example, consolidations are small leading to many
 false-negatives. 2) Low 
inter-class variance. GGO boundaries
often have low contrast and blurred appearances making their identification a challenge for algorithms. 3) Difficulties in collecting a large labeled database within a a short time frame for DL systems.

\begin{figure*}[t]
 \centering
\begin{tabular}{ccccc}
\includegraphics[height=3cm, width=3cm]{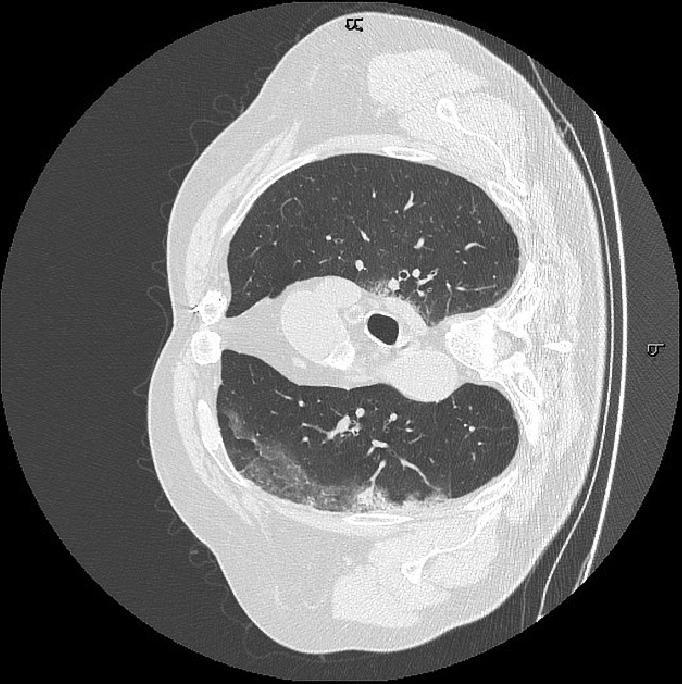} &
\includegraphics[height=3cm, width=3cm]{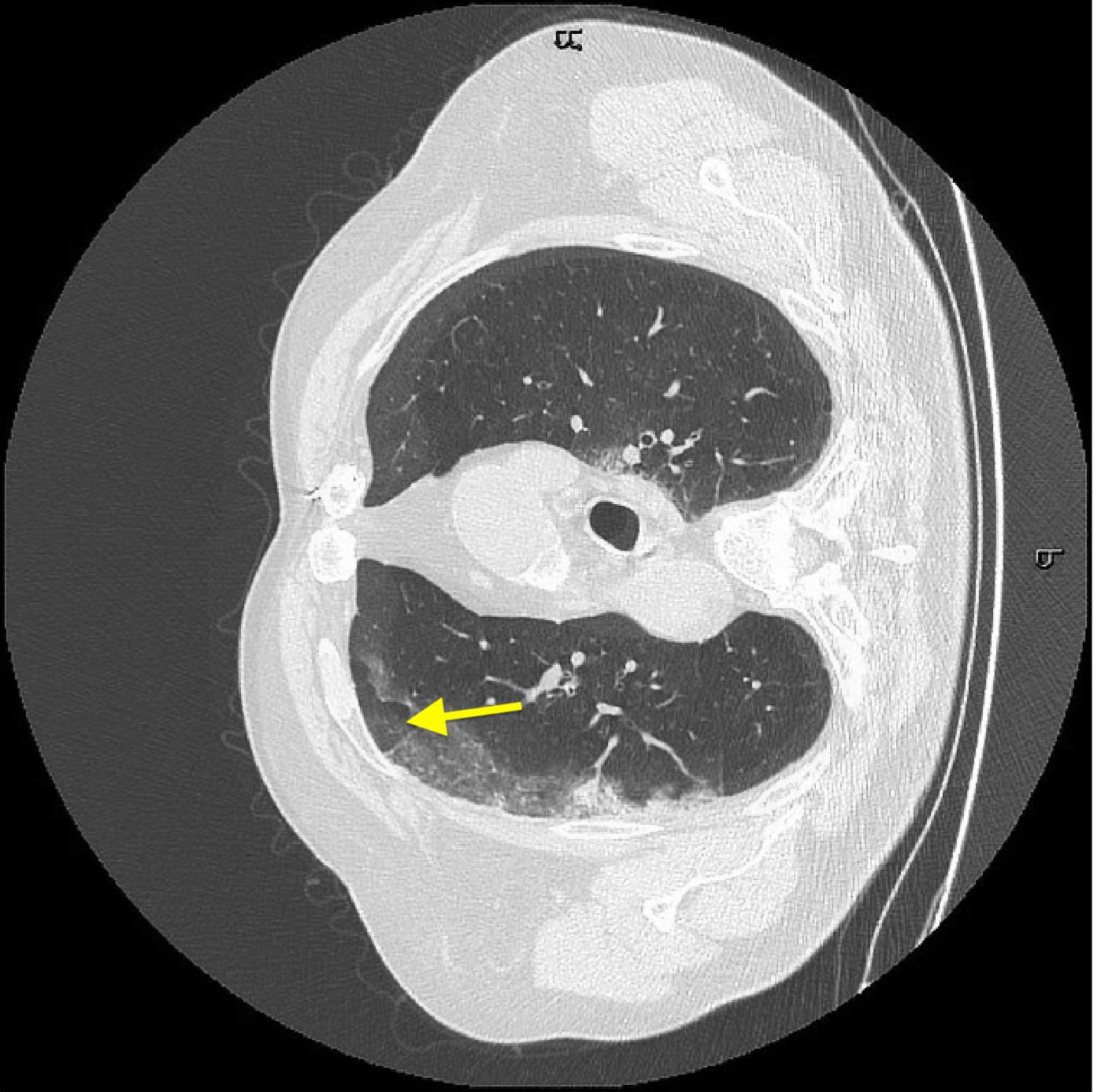} &
\includegraphics[height=3cm, width=3cm]{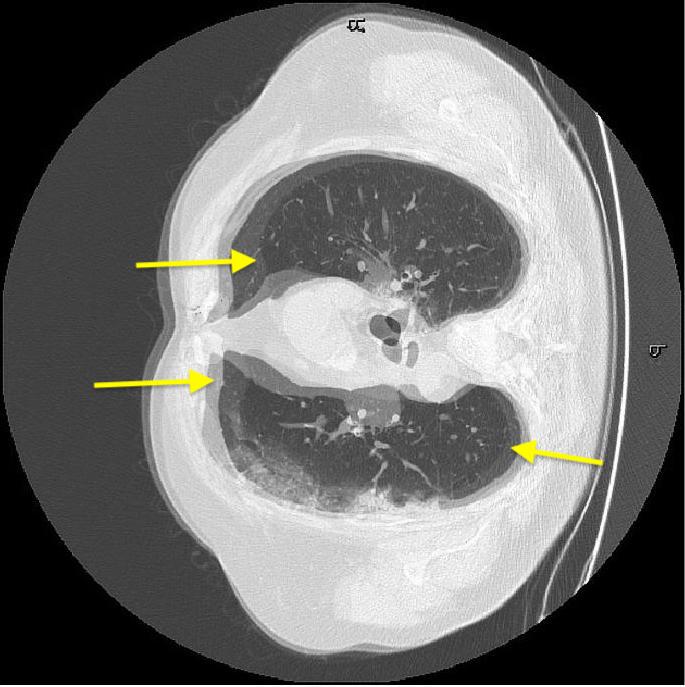} &
\includegraphics[height=3cm, width=3cm]{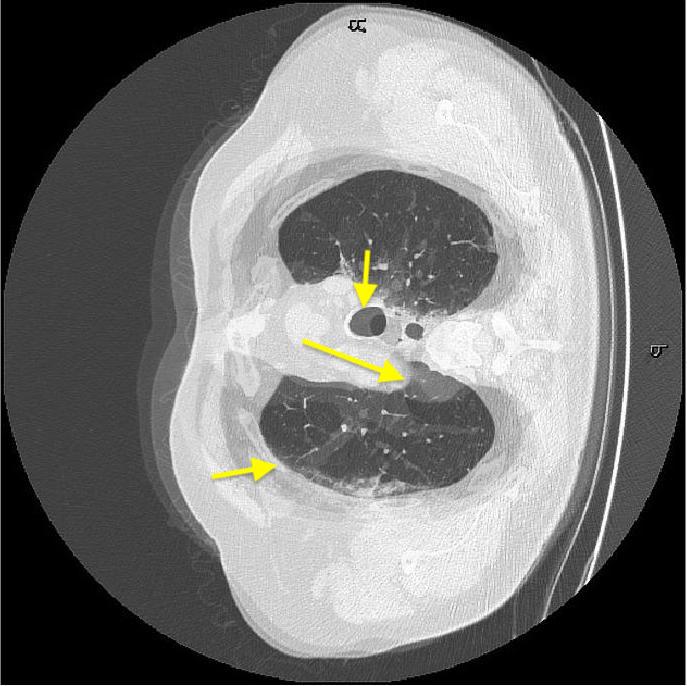} &
\includegraphics[height=3cm, width=3cm]{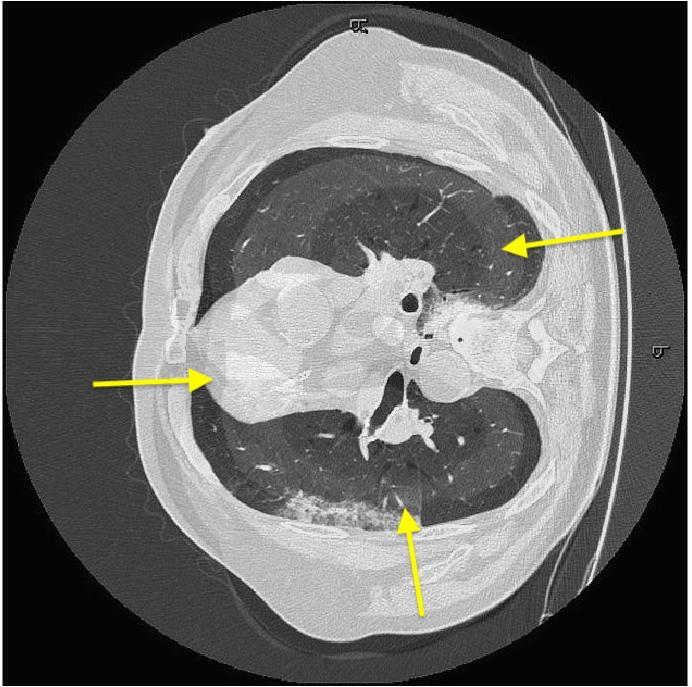} \\
(a) & (b) & (c) & (d) & (e) \\
\end{tabular}
\caption{(a) Base image with lung and infection regions highlighted; Example of generated images using: (b) Our proposed $GeoGAN$ method; (c) \cite{Zhao_CVPR2019}; (d) $DAGAN$ method by \cite{DAGAN}; (e) $cGAN$ method by \cite{Mahapatra_MICCAI2018}. Artifact regions are indicated by yellow arrows. }
\label{fig:SynImages_comp}
\end{figure*}

Furthermore, acquiring high quality
pixel-level annotation of lung infections in CT scans
is expensive and time-consuming. 
Manual delineation of lung infections is tedious
and time-consuming, and infection annotations is a highly subjective task.
 Large scale dataset annotations for segmentation require pixel labels, which is time consuming and involves high degree of clinical expertise. The problem is particularly acute for pathological images (as in the case of COVID infections) since it is difficult to obtain diverse images for less prevalent disease conditions, necessitating data augmentation. We propose a generative adversarial network (GAN) based approach for pathological image augmentation and demonstrate its efficacy in COVID pathological region segmentation. Figure~\ref{fig:SynImages_comp} shows example cases of synthetic images generated by our method and other competing techniques. %Yellow arrows show image locations with prominent artifacts, and such training images will not lead to robust diagnosis systems. %summarizes the image generation results of our approach and \cite{Zhao_CVPR2019}, and highlights our superior performance by incorporating geometric information.

Traditional augmentations such as image rotations or deformations have limited benefit as they do not fully represent the underlying data distribution of the training set and are sensitive to parameter choices. %Recent works \cite{huang2018auggan,Zhao_CVPR2019, han2018gan, nielsen2019gan} proposed to solve this issue by using synthetic data for augmentation and increase diversity in the training samples. However, certain challenges have not been satisfactorily addressed by these methods. 
%
%  Zhao et. al. \cite{Zhao_CVPR2019} proposed a learning-based registration method to register images to an atlas, use  corresponding deformation field to deform a segmentation mask and obtain new data. This approach presents the following challenges: 1) since registration errors propagate to subsequent stages, inaccurate registration can adversely affect the data generation process; 2) with atlas of a normal subject it is challenging to register images from diseased subjects due to appearance or shape changes. This is particularly relevant for layer segmentation in retinal optical coherence tomography (OCT) images, where there is a drastic difference in layer shape between normal and diseased cases. Figure~\ref{fig:OCT1} (a) shows the retinal layers of a normal subject, and Figure~\ref{fig:OCT1} (b) shows two cases of retinal fluid build up due to diabetic macular edema (DME) and age related macular degeneration (AMD). The retinal layers are severely distorted compared to Figure~\ref{fig:OCT1} (a) and registration approaches have limited impact in generating accurate images. 
%
Recent data augmentation methods of \cite{han2018gan, nielsen2019gan,Mahapatra_CVIU19,MahapatraISBI08,MahapatraICME08,MahapatraICBME08_Retrieve,MahapatraICBME08_Sal,MahapatraSPIE08,MahapatraICIT06,CVPR2020_Ar,sZoom_Ar,Kuanar_AR1,Lie_AR2,Lie_AR,Salad_AR,Stain_AR,DART2020_Ar,TMI2021_Ar} use generative adversarial network (GAN), [\cite{goodfellow2014generative}], and show moderate success for medical image classification. However, they have limited relevance for segmentation since they do not model geometric relation between different organs and most augmentation approaches do not differentiate between normal and diseased samples.  %
%
% 
% Experiments in Section~\ref{expt:dis} show segmentation methods trained on normal subject images are not  equally effective for diseased cases due to significant shape changes between the two types. 
Hence there is a need for augmentation methods that consider the geometric relation between different anatomical regions and generate distinct images for diseased and normal cases.
%
% Another limitation of current augmentation approaches is that they do not incorporate diversity in a principled manner. In our proposed method we  \cite{Mahapatra_MICCAI2018} shape mask was incorporated manually for image generation, which is not practical and may lead to unrealistic deformations. 

\section{Related Work}
\label{sec:prior}

\subsection{Chest CT Segmentation}

Segmentation of lungs from chest CT scans is a widely explored topic \cite{InfNet_22,CVIU_Ar,AMD_OCT,GANReg1_Ar,PGAN_Ar,Haze_Ar,Xr_Ar} since it facilitates  diagnosis and quantification of  lung diseases \cite{InfNet_24}. \cite{InfNet_26,RegGan_Ar,ISR_Ar,LME_Ar,Misc,Health_p,Pat2,Pat3,Mahapatra_Thesis} use support
vector machines (SVM) to detect lung nodules
from CT scans.% while \cite{InfNet_27} use bidirectional chain codes. 
 Nodule extraction is challenging due to similar appearance with the background. Deep learning algorithms have been able to overcome this challenge by learning powerful discriminative features. \cite{InfNet_28,Pat4,Pat5,Pat6,Pat7,Pat8,Pat9} use CNNs to segment lung nodules from heterogeneous CT scans. \cite{InfNet_29,Pat10,Pat11,Pat12,Pat13,Pat14,Pat15,Pat16,Pat17,Pat18} make use of GAN-synthesized data to improve the performance of a discriminative model for pathological lung segmentation. \cite{InfNet_30} employ two deep networks to segment lung tumors from CT scans by adding multiple residual streams of varying resolutions.

\subsection{Deep Learning For Imaging Based COVID-19 Analysis}

% Artificial intelligence technologies have been employed in
% a large number of applications against COVID-19 [6], [15].
% Joseph et al. [15] categorized these applications into three
% scales, including patient scale (e.g., medical imaging for
% diagnosis [36], [37]), molecular scale (e.g., protein structure
% prediction [38]), and societal scale (e.g., epidemiology [39]).
% In this work, we focus on patient scale applications [18], [21],
% [36], [37], [40]–[43], especially those based on CT scans.

\cite{InfNet_36} use a modified inception network of \cite{InfNet_44} for classifying COVID-19 patients and
normal controls. Instead of directly training on complete CT
images, they trained the network on the regions of interest,
which are identified by two radiologists based on the features of pneumonia. \cite{InfNet_37} use CT images to train a U-Net++ \cite{InfNet_45} for identifying COVID-19 patients that performs comparably with expert radiologists. DL approaches have also been used for segmenting infection regions in lung CT \cite{InfNet} and for lung infection quantification \cite{InfNet_15,InfNet_20,InfNet_21} of COVID-19.

\subsection{Data Augmentation (DA)}
While conventional augmentation approaches (such as rotation, scaling, etc) can generate a large database, they do not add much data diversity. They are also sensitive to parameter values \cite{Zhao19_25}, variation in image resolution, appearance and quality \cite{Zhao19_45}.
 Recent DL based methods trained with synthetic images outperform those trained with standard DA over classification and segmentation tasks. \cite{DAGAN} proposed DAGAN for image generation in  few shot learning systems. \cite{Zhao_CVPR2019} proposed a learning-based registration method to register images to an atlas, use  corresponding deformation field to deform a segmentation mask and obtain new data. %\cite{bozorgtabar2019syndemo} used GAN objective for domain transformation by aligning feature distribution of target data  and source domain.
 \cite{Mahapatra_MICCAI2018} used conditional GAN (cGAN) for generating informative synthetic chest Xray images conditioned on a perturbed input mask.

%
%
% GANs have also been used for generating synthetic retinal images \cite{ZhaoMIA2018} and brain magnetic resonance images (MRI) \cite{han2018gan,ShinSASHIMI2018}, facial expression analysis \cite{Behzad_PR2020}, for super resolution \cite{SRGAN,MahapatraMICCAI_ISR}, %image registration \cite{Mahapatra_PR2020,Mahapatra_ISBI19,Mahapatra_MLMI18} 
% and generating higher strength MRI from their low strength acquisition counterparts \cite{GAN_MI_Rev}. 
GANs have also been used for generating synthetic retinal images in \cite{ZhaoMIA2018} and brain magnetic resonance images (MRI) in \cite{han2018gan,ShinSASHIMI2018}, image registration \cite{Mahapatra_PR2020} 
and generating higher strength MRI from their low strength acquisition counterparts \cite{GAN_MI_Rev}. 
 Generated images have implicit variations in intensity distribution but there is no explicit attempt to model attributes such as shape variations that are important to capture different conditions across a population. 
 \cite{Zhao19_51} augmented medical images with simulated anatomical variations but demonstrate inconsistent performance based on transformation functions and parameter settings.
%------------------------------------

\subsection{Image Generation Using Uncertainty}
 
 \cite{PhS2} used approximate Bayesian inference for parameter uncertainty estimation in scene understanding, but did not capture complex correlations between different labels. \cite{PhS5} proposed a method to generate different samples using an ensemble of $M$ networks while \cite{PhS7} present a single network with $M$ heads for image generation. \cite{PhS8} proposed a method based on conditional variational autoencoders (cVAE) to model segmentation masks, which improves the quality of generated images. In probabilistic UNet \cite{PhS4}, cVAE is combined with UNet \cite{Unet} to generate multiple segmentation masks, although with limited diversity since randomness is introduced at highest resolution only. \cite{PhiSeg} introduced a framework to generate images with a greater diversity by injecting randomness at multiple levels.

\subsection{Our Contribution}

Since annotating medical images is a time consuming task, it is challenging to obtain manually annotated segmentation masks to model the geometrical relation between different labels in the image. To overcome this challenge we propose to use a weakly supervised segmentation approach to generate labeled segmentation maps. The generated segmentation maps are then used to model the geometric relationship between the different pathological regions.

Based on the premise that improved data augmentation yields better segmentation performance in a DL system, we hypothesize that improved generation of synthetic images is possible  by considering the intrinsic relationships between shape and geometry of anatomical structures \cite{ShapeSim}.  In this paper we present a Geometry-Aware Shape Generative Adversarial Network (GeoGAN)\footnote{A pre-print of a preliminary version of our method applied to fluid segmentation from retinal OCT scans can be found at https://arxiv.org/pdf/2003.14119.pdf. We introduce an additional weakly supervised segmentation step. Since the current submission is for a COVID special issue the results from the pre-print are not included.} that learns to generate plausible images of the desired anatomy (e.g.,COVID infected areas in the lung) while preserving learned relationships between geometry and shape. We make the following contributions: 
 \begin{enumerate}
     \item Incorporating geometry information contributes to generation of realistic and qualitatively different medical images and \textbf{shapes} compared to standard DA. Other works such as \cite{Mahapatra_MICCAI2018,ZhaoMIA2018} do not incorporate this geometric relationship between anatomical parts.
    
     \item Use of uncertainty sampling and conditional shape generation on class labels to introduce diversity in the mask generation process. Compared to previous methods  we introduce diversity at different stages (different from \cite{Mahapatra_MICCAI2018,ZhaoMIA2018,PhS4}) and introduce an auxiliary classifier (different from \cite{PhiSeg,PhS8} ) for improving the quality and accuracy of generated images.
 \end{enumerate}

\section{Method}
\label{sec:method}

Our augmentation method: 1) models geometric relationship between multiple segmentation labels; 2) preserves disease class label of original image to learn disease specific appearance and shape characteristics; and 3) introduces diversity in the image generation process through uncertainty sampling.
We demonstrate our method's capability by training it on a dataset of CT lung images having annotations of COVID infected areas. However, in real world scenarios it is difficult to find datasets with such detailed annotations, especially in the case of COVID-19. Hence we introduce a weakly supervised segmentation (WSS) step that segments a CT image into different labeled regions using only the image labels of prevalent pathologies. The resulting label map enables us to learn the geometric relationship between different labels, which is essential to synthesize realistic images for data augmentation. 

\begin{figure*}[h]
 \centering
\begin{tabular}{c}
\includegraphics[height=8cm, width=15cm]{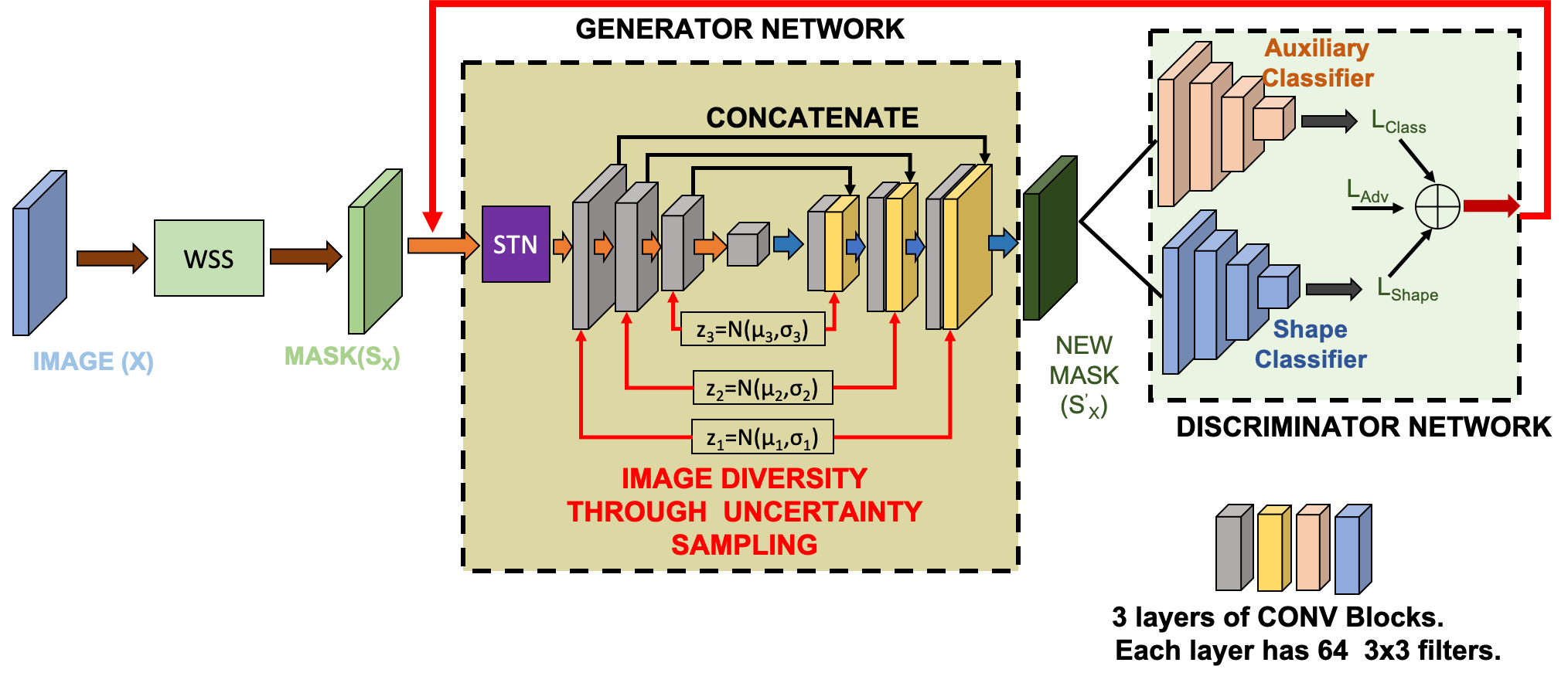}  \\
\end{tabular}
\caption{Overview of the steps in the training stage of our method. The images ($X$) and corresponding segmentation masks ($S_X$) are input to a STN whose output is fed to the generator network. Generator network is based on UNet architecture, and diversity through uncertainty sampling is injected at different levels. The generated mask $S_X^{'}$ is fed to the discriminator which evaluates its accuracy based on $L_{class}$, $L_{shape}$ and $L_{adv}$. The provided feedback is used for weight updates to obtain the final model.}
\label{fig:workflow}
\end{figure*}

\begin{figure}[h]
 \centering
\begin{tabular}{c}
\includegraphics[height=3.1cm, width=7.9cm]{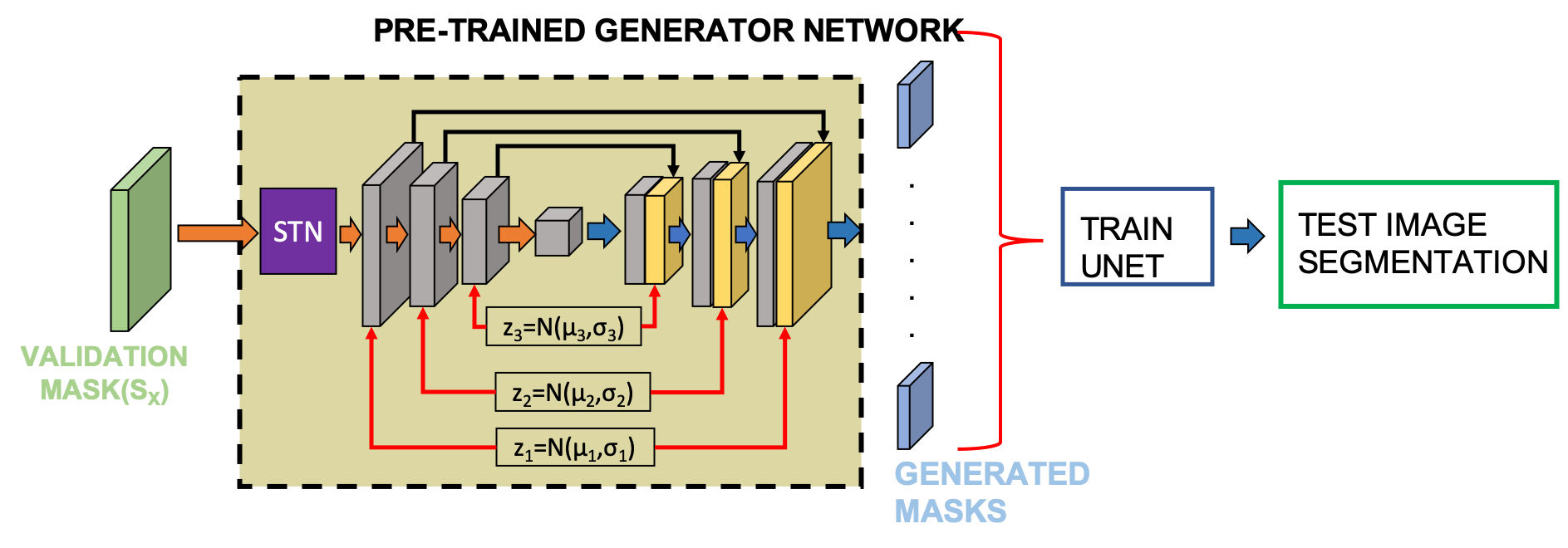}  \\
\end{tabular}
\caption{Depiction of mask generation. The trained generator network is used on validation set base images and masks to generate new images that are used to train a segmentation network (UNet or Dense UNet). The model then segments infected regions from test images.}
\label{fig:workflow2}
\end{figure}

Figure~\ref{fig:workflow} shows the training workflow using a modified UNet based generator network. The set of images and their WSS-obtained segmentation masks are used to train the generator while the discriminator provides feedback to improve the generator output. Figure~\ref{fig:workflow2} depicts generation of synthetic images after training is complete and their subsequent use in training a UNet for image segmentation at test time.

\subsection{Weakly Supervised Segmentation}
\label{met:wss}

In order to obtain pixel labels from the image labels in a weakly supervised setting we solve a instance-level classification problem where pixels are instances. Subsequently, existing well-developed fully supervised segmentation methods can be applied.
We use a combined Multiple Instance Learning (cMIL) for instance classification \cite{CAMEL}. The image is split into $N \times N$ grids (instances) of equal size where instances from the same image are in the same bag. In
cMIL, two MIL-based classifiers with different instance selection criteria ($Max-Max$ and $Max-Min$) are used to select instances to construct the instance-level dataset for subsequent classification.

\begin{figure}[h]
 \centering
\begin{tabular}{cc}
\includegraphics[height=2.5cm, width=3.5cm]{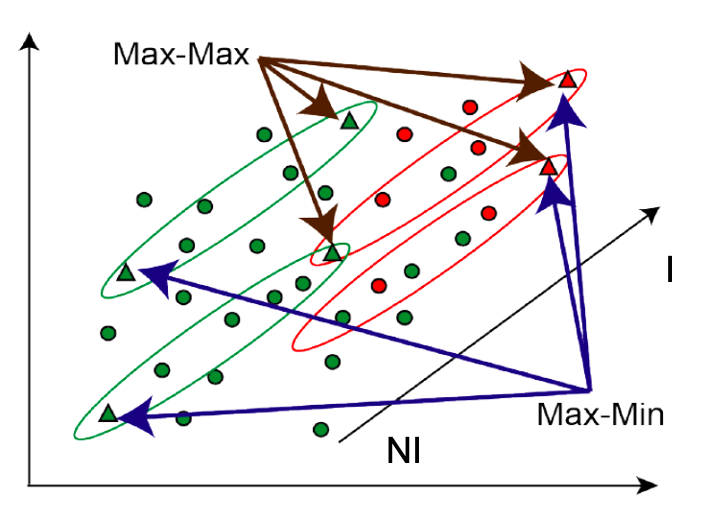}  & 
\includegraphics[height=2.5cm, width=3.5cm]{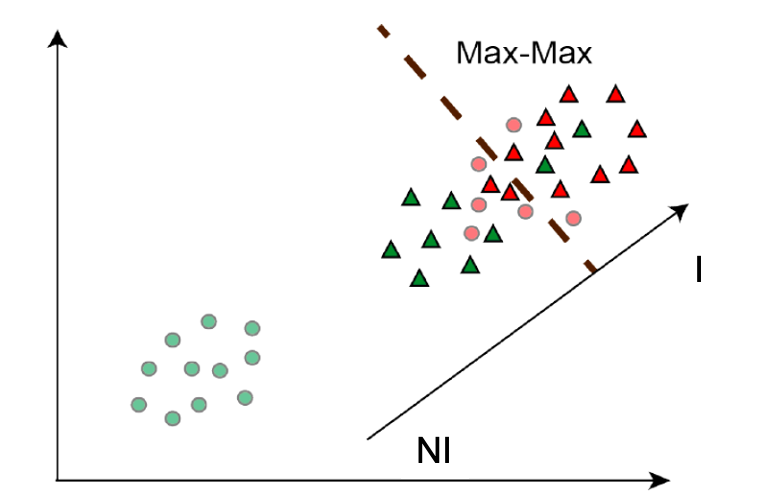}  \\
(a) & (b) \\
\includegraphics[height=2.5cm, width=3.5cm]{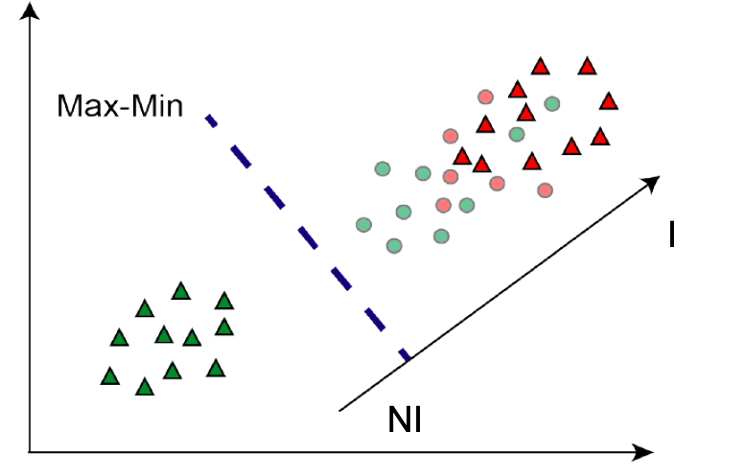}  & 
\includegraphics[height=2.5cm, width=3.5cm]{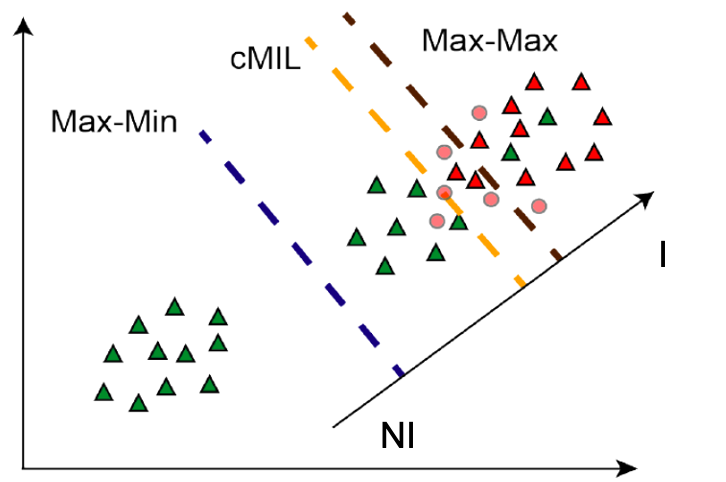}  \\
(c) & (d) \\
\end{tabular}
\caption{Intuition behind two instance selection criteria named $Max-Max$ and $Max-Min$. Red and green circles represent the $I$ and $NI$ instances, respectively. We use triangles to represent the selected instances, and circles with light colors to represent the instances that are not selected. Each dotted line represents the decision boundary of the classifier, which is trained with the selected instances. Each ellipse represents an image (or a bag in MIL). cMIL, which combines $Max-Max$ and $Max-Min$, achieves a better decision boundary.}
\label{fig:MaxMin}
\end{figure}

The selected instance can be considered as the representative of its corresponding image, which determines the image
class.% (similar to the attention mechanism in \cite{CAMEL_24}).
If the image is labeled `infected' ($I$) we  reason that at least one instance is infected. On the other
hand, if the label of the image is `not infected' ($NI$), all the instances in it are $NI$. For both $I$ and $NI$
images, $Max-Max$ selects the instance with maximum $I$
response. As shown in Figures~\ref{fig:MaxMin} (a) and (b), during the training stage the $Max-Max$ criterion will select the
instance with maximum $I$ response as the $NI$ example from the $NI$ samples. Therefore, the
model trained with these data would give a decision boundary biased towards $I$ leading to misclassification of $I$ instances with lower responses (as shown by light red circles). For example, $I$ instances with similar appearances to $NI$ may get misclassified.

$Max-Min$ acts as a countermeasure that selects the instances
with the highest $I$ response for $I$ images and the instances
with the lowest response for $NI$ images. As shown
in Figure~\ref{fig:MaxMin}(c), $Max-Min$ tends to have an opposite effect compared
to $Max-Max$. Therefore, in cMIL we combine these
two criteria to obtain a  balanced instance-level dataset to be used in
fully supervised learning (see Figure\ref{fig:MaxMin} (d)). It is worth noting
that, for $NI$ images, although each instance is $NI$, we only use the selected instances to avoid data imbalance.

Using a ResNet-50 architecture we train the two MIL-based classifiers separately under the same configuration: in the forward pass, we use the $Max-Max$ (or $Max-Min$ for the other classifier) criterion to select
one instance from each bag based on their predictions, and the prediction of the selected instance is regarded as the prediction
of the image. In the backpropagation step, we use the cross entropy
loss between the image-level label and the prediction
of the selected instance to update the classifier’s parameters.
The loss function for each classifier is defined as follows:
\begin{equation}
    Loss= -\sum_j \left(y_j\log\hat{p}_j + (1-y_j)\log(1-\hat{p}_j)  \right),
\end{equation}
where $\hat{p}_j=S_{criterion}(\{f(b_i)\})$, $b_i$ are the instances in image $j$, $f$ is the classifier, $S_{criterion}\in \{Max-Max,Max-Min\}$.
$S_{criterion}$ selects the target instance using the defined criterion, $y_j$ is the image-level label.

For Max-Max criterion:
\begin{equation}
    S_{Max-Max}(\{f(b_i)\}) = \max_i {f(b_i)}
\end{equation}
   
For Max-Min criterion:
\begin{equation}
S_{Max-Min}({f(b_i)}) =
\begin{cases}
\max_i {f(b_i)} & \text{if} ~y = 1 \\
\min_i {f(b_i)} & \text{if} ~y = 0
\end{cases}
\end{equation}

After training, we feed the same training data into
the two trained classifiers and select the instances under the corresponding criterion, then the predictions are considered as their labels. We combine the instances selected by the two trained classifiers to construct the final fully supervised
instance-level dataset. Note that we discard those potentially confusing samples whose predicted labels are different from their corresponding image-level labels.

\paragraph{Retrain and Relabel}
Once the instance-level dataset is selected, we 
train an instance classifier in a fully supervised manner. Similar to cMIL we use a ResNet-50 and name this step as \textit{retrain}. Then, we split the original image into latticed instances
and relabel them using the trained instance-level
classification model. For each image, we obtain
$N^{2}$ high-quality instance labels from a single image-level label.

\subsubsection{Segmentation}

With enriched supervision information, the instance level
labels are directly assigned to the corresponding pixels,
producing approximate pixel-level labels. Therefore,
we can train segmentation models in a fully supervised way using well-developed architectures such as UNet++ \cite{InfNet_45}. 

% To prevent the model from learning the
% checkboard-like artifacts in the approximate labels, in the
% training process, we perform data augmentation by feeding
% smaller images that are randomly cropped from the original
% training set and their corresponding masks into the segmentation
% model.

% 3.3. Further Improvement
% Value of $N$ plays an important role in the quality of labels. Higher $N$ leads to  finer labels but lead
% to severe image information loss. We tackle this issue in the following steps

% At each instance selection criterion only choose one instance
% from the image to construct the instance-level dataset,
% which only takes up a small portion of the image, resulting
% in losing a considerable amount of image information from
% the original image-level dataset. In order to recover this information
% loss and increase data diversity in the instancelevel
% dataset, we further introduce the cascade data enhancement
% method to generate the instance-level dataset by
% two concurrent routes (Fig. 4). Here, we use cMIL(N) to
% denote the cMIL with a scale factor of N. To derive labeled
% instances of a scale factor of N, we can either use
% cMIL(N) or cMIL(N1) and cMIL(N2) back-to-back where
% N = N1 × N2. The two sources of data are combined before
% fed into the segmentation model.

\paragraph{ Training with Image-Level Constraints}

In order to maximize the utility of the original image-level
supervision information, in the retrain step, we can
add the original image-level data as one additional input
source going through the classifier. The image-level constraint is imposed under Max-Max and
Max-Min criteria to the instance level, the total loss is defined
as the sum of the retrain loss and the constraint loss:
\begin{equation}
Loss = w_1\times Loss_{constrain} + w_2\times Loss_{retrain} 
\end{equation}
where $w_1$ and $w_2$ are the weights of the two losses. We set $w_1 = w_2$ in our experiments.
\begin{equation}
Loss_{constrain} = -\sum_{S_{criterion}} (y \log \hat{p} + (1-y)\log(1-\hat{p})),
\end{equation}
where $\hat{p} = S_{criterion}({f(b_i)})$, $b_i$ represents the selected instance, $f$ is the image-level constrain route, $S_{criterion} \in
\{Max-Max, Max-Min\}$, and $y$ is the image-level label.
\begin{equation}
Loss_{retrain} = -\sum_j (y_j \log \hat{y}_j+(1-y_j) log(1-\hat{y}_j))
\end{equation}
where $\hat{y}_j = g(n_j)$, $n_j$ represents the input instance, $g$ is the retrain route, and $y_j$ is the instance-level label. %Since two routes share the same network, we have f  g.

\subsection{Geometry Aware Shape Generation}

Let us denote an input image as $x$, the corresponding  segmentation masks as $s_x$ and the disease class label of $x$ as $l_x$. Our method learns to generate a new image and segmentation label map from a base image and its corresponding mask.  
  The first stage is a spatial transformer network (STN) \cite{STN} that transforms the base mask to a new shape with different attributes of location, scale and orientation. 
 The transformations used to obtain new segmentation mask $s_x^{'}$ are applied to $x$ to get corresponding transformed image $x^{'}$.
Since the primary aim of our approach is to learn contours and other shape specific information of anatomical regions, a modified UNet architecture as the generator network effectively captures hierarchical information of shapes. It also makes it easier to introduce diversity at different levels of image abstraction.

The generator $\textbf{G}_g$ takes input $\textbf{s}_x$ and a desired label vector of output mask $c_g$ to output an affine transformation matrix $\textbf{A}$ via a STN, i.e., $\textbf{G}_g$($\textbf{s}_x, c_g) = \textbf{A}$.  
$\textbf{A}$ is used to generate  $s_x^{'}$  and  $x^{'}$. The discriminator $\textbf{D}_{class}$ determines whether output image preserves the desired label $c_g$ or not. 
The discriminator $\textbf{D}_g$ is tasked with ensuring that the generated masks and images are realistic. Let the minimax criteria between $\textbf{G}_g$ and $\textbf{D}_g$ be $\min_{\textbf{G}_g} \max_{\textbf{D}_g} \textbf{L}_g(\textbf{G}_g,\textbf{D}_g)$. The loss function $\textbf{L}_g$ has three components
 \begin{equation}
     L_g=L_{adv} + \lambda_1{L}_{class} + \lambda_2{L}_{shape}
 \label{eqn:Tloss}
 \end{equation}
 where 1) $\textbf{L}_{adv}$ is an adversarial loss to ensure $\textbf{G}_{g}$ outputs realistic deformations; 2) $\textbf{L}_{class}$ ensures generated image has characteristics of the target output class label (disease or normal); and 3) $\textbf{L}_{shape}$ ensures new masks have realistic shapes. $\lambda_1,\lambda_2$ balance each term's contribution. 

\paragraph{\textbf{Adversarial loss}}- $\textbf{L}_{adv}(\textbf{G}_g,\textbf{D}_g)$: The STN outputs $\widetilde{A}$, a prediction for $\textbf{A}$ conditioned on $\textbf{s}_x$ and a new semantic map $\textbf{s}_x \oplus \widetilde{A}(\textbf{s}_x)$ is generated. % using  composing a transformed box onto the input.
 $\textbf{L}_{adv}$ is  defined as:
% %
\begin{equation}
\begin{split}
L_{adv}(G_g,D_g) =\mathbb{E}_x\left[\log D_{g} (\textbf{s}_x \oplus \widetilde{A}(\textbf{s}_x)) \right] \\
+ \mathbb{E}_{\textbf{s}_x} \left[\log (1-D_{g} (\textbf{s}_x \oplus \widetilde{A}(\textbf{s}_x))) \right] ,
\end{split}
\label{eq:D2}
\end{equation}
% %

\paragraph{Classification Loss}- $\textbf{L}_{class}$: The affine transformation $\textbf{A}$ is applied to the base image $\textbf{x}$ to obtain  the generated image $\textbf{x}^{'}$. We add an auxiliary classifier when optimizing both $\textbf{G}_g$ and $\textbf{D}_g$ and define the classification loss as,
\begin{equation}
L_{class} = \mathbb{E}_{\textbf{x}^{'},c_g} [-\log D_{class}(c_g|x')],
\label{eq:sgan2}
\end{equation}
where the term $D_{class}(c_g|x')$ represents a probability distribution over classification labels computed by $D$. 

\paragraph{Shape Loss}-$\textbf{L}_{shape}$:
We intend to preserve the relative geometric arrangement between the different labels. 
The generated mask has regions with different assigned segmentation labels because the base mask (from which the image was generated) already has labeled layers. Let us denote by $s_i$ the image region (or pixels) in $\textbf{s}_x$ assigned label $i$. Consider another set of pixels, $s_j$, assigned label $j$. We calculate $P_{shape}(l_i|s_j,s_i)$, which is,  given regions $s_i,s_j$, the pairwise probability of $s_i$ being label $i$. 
If $n$ denotes the  total number of labels, for every label $i$ we calculate the $n-1$ such probability values and repeat it for all $n$ labels. Thus 
\begin{equation}
L_{shape} = \frac{1}{n\times(n-1)} \sum_{i,j}^{i\neq j} P_{shape} ; ~ (i,j) \in \{1,\cdots,n\}
\label{eq:sgan3}
\end{equation}
The probability value is determined from a pre-trained modified VGG16 architecture to compute $L_{shape}$ where the input has two separate maps corresponding to the label pair. Each map's foreground has only the region of the corresponding label and other labels considered background.  The conditional probability between the pair of label maps enables the classifier to implicitly capture geometrical relationships and volume information between the label pair without the need to define explicit features. The geometric relation between different labels will vary for infected and non-infected cases, which is effectively captured by our approach.
To get the pre-trained VGG16 network we used a separate dataset of $24$ images with its WSS generated segmentation maps. %manually annotated images. %In Section~\ref{expt:ablation2} we analyze the effect of using manually annotated images versus using the segmentation maps obtained from the WSS step.

% Since we are proposing a weakly supervised segmentation based approach and work with the assumption that we do not have manually labeled masks, we 
%  

\subsection{Sample Diversity From Uncertainty Sampling}

The generated mask $s_x'$ is obtained by fusing $L$ levels of the generator $G_g$ (as shown in Figure~\ref{fig:workflow}), each of which is associated with a latent variable $z_l$. We use probabilistic uncertainty sampling to model conditional distribution of segmentation masks and use separate latent variables at multi-resolutions to factor inherent uncertainties.
The hierarchical approach introduces diversity at different stages and influences different features (e.g., low level features at the early layers and abstract features in the later layers). Denoting the generated mask as $s$ for simplicity, we obtain conditional distribution $p(s|x)$ for $L$ latent levels as:
\begin{equation}
\begin{split}
p(s|x) = \int p(s|z_1 , \cdots, z_L )p(z_1 |z_2 , x) \cdots \\ 
     p(z_{L-1} |z_L , x)p(z_L |x) dz_1 \cdots dz_L . 
\end{split}
\label{eq:prob1}
\end{equation}

Latent variable $z_l$  models diversity at resolution  $2^{-l+1}$ of the original image (e.g. $z_1$ and $z_3$ denote the original and $1/4$ image resolution). 
A variational approximation $q(z|s, x)$ approximates the posterior distribution $p(z|s,x)$  where $z=\{z_1 , . . . , z_L\}$. $ \log p(s|x) = L(s|x) + KL(q(z|s, x)||p(z|s, x))$, where $L$ is the evidence lower bound, and $KL(.,.)$ is the Kullback-Leibler divergence. %Since $KL(·, ·) \geq 0$, $L$ is a lower bound on the conditional log probability when the approximation $q$ exactly matches the posterior. 
 The prior and posterior distributions are parameterized as normal distributions $\mathcal{N}(z|\mu,\sigma)$. %Thus, we define

Figure~\ref{fig:workflow}  shows example implementation for $L=3$. We use $6$ resolution levels and $L = 4$ latent levels.
Figure~\ref{fig:workflow}  shows the latent variables $z_l$ forming skip connections in a UNet architecture such that information between the image and segmentation output goes through a sampling step. The latent variables \emph{are not mapped} to a 1-D vector to preserve the structural  relationship between them, and this substantially improves segmentation accuracy. 
$z_l$'s dimensionality is $r_x 2^{-l+1} \times r_y 2^{ -l+1}$, where $r_x$ , $r_y$ are image dimensions. %$z_l$  models image data at $2^{-l+1}$ of original resolution due to downsampling operations and transmits the learned representation to the latent space embedding above ($z_{l-1}$). 

%
%

% this file describes the experiments and results for the paper

\section{Experimental Results}
\label{sec:expt}

\subsection{Dataset Description}
 We use the following three different segmentation datasets\footnote{http://medicalsegmentation.com/covid19/}:
 
\textbf{CT Segmentation Dataset 1 (CTSeg1)}:
 The dataset  consists of 100 axial CT images from different COVID-19 patients. All the CT images were collected by the Italian Society of Medical and Interventional Radiology. A radiologist segmented the CT images with $3$ labels: ground-glass (mask value =1), consolidation (=2) and pleural effusion (=3). 

\textbf{CT Segmentation Dataset 2 (CTSeg2)}:
The second dataset is collected from $9$ axial volumetric CTs\footnote{https://radiopaedia.org/articles/covid-19-3}. It includes whole volumes and both positive and negative slices ($373$ out of the total of $829$ slices have been evaluated by a radiologist as positive and segmented). %These volumes are converted and normalized in a similar way as $CTSeg1$. 

\textbf{CT Segmentation Dataset 3 (CTSeg3)}:
This dataset contains $20$ labeled COVID-19 CT scans\footnote{https://zenodo.org/record/3757476.XsqZCJ4zadZ}. Left lung, right lung, and infections are labeled by two radiologists and verified by an experienced radiologist.

% In this work, we collected a semi-supervised COVID-19
% infection segmentation dataset (COVID-SemiSeg), to leverage
% large-scale unlabeled CT images for augmenting the training
% dataset. We employ COVID-19 CT Segmentation [9] as the labeled
% data DLabeled, which consists of 45 CT images randomly
% selected as training samples, 5 CT images for validation, and
% the remaining 50 images for testing. The unlabeled CT images
% are extracted from the COVID-19 CT Collection [11] dataset,
% which consists of 20 CT volumes from different COVID-
% 19 patients. We extracted 1,600 2D CT axial slices from
% the 3D volumes, removed non-lung regions, and constructed
% an unlabeled training dataset DUnlabeled for effective semisupervised
% segmentation.

\subsection{Experimental Setup, Baselines and Metrics}

Our method  has the following steps: 1) Use the default training, validation, and test folds of the dataset. 2) Use training images to train the image generator. 3) Generate shapes from the training set and train UNet++ segmentation network \cite{InfNet_45} on the generated images. 4) Use trained UNet++ to segment test images. 5) Repeat the above steps for different data augmentation methods. 
%a NVIDIA Titan X GPU having $12$ GB RAM, 
Our model is implemented in PyTorch, on a NVIDIA TITAN X GPU.
 We trained all models using Adam optimiser \cite{Adam} with a learning rate of $10^{-3}$ and batch-size of $16$. Batch-normalisation was used. % .
 The values of parameters $\lambda_1$ and $\lambda_2$ in Eqn.~\ref{eqn:Tloss} were set by a detailed grid search on a separate dataset of $14$ volumes that was not used for training or testing. They were varied between $[0,1]$ in steps of $0.05$ by fixing $\lambda_1$ and varying $\lambda_2$ for the whole range. This was repeated for all values of $\lambda_1$. The best segmentation accuracy was obtained for $\lambda_1=0.92$ and $\lambda_2=0.9$, which were our final parameter values.

We denote our method as  $GeoGAN_{WSS}$ (Geometry Aware GANs using the weakly supervised segmentation component), and compare it's performance against other methods such as: 1) rotation, translation and scaling (denoted as DA-Data Augmentation); 2) $DAGAN$ - data augmentation GANs of \cite{DAGAN}; 3) $cGAN$ - the conditional GAN based method of \cite{Mahapatra_MICCAI2018}; 4) $Zhao$- the atlas  registration method of \cite{Zhao_CVPR2019}; 5) $GeoGAN_{Manual}$ - Geometry Aware GANs using the manual segmentation maps for image synthesis.
%
%  %
Segmentation performance is evaluated in terms of Dice Metric (DM), Hausdorff Distance (HD) and Mean Absolute error (MAE). DM of $1$ indicates perfect overlap and $0$ indicates no overlap, while lower values of MAE indicate better segmentation performance.

%  For the classification dataset we use area under curve(AUC) and accuracy as metrics , while for segmentation we report the Dice Metric (DM) and Hausdorff distance (HD) values.
 
\paragraph{Algorithm Baselines.}

The following variants of our method were used for ablation studies: 
\begin{enumerate}
    \item  GeoGAN$_{noL_{class}}$- GeoGAN$_{WSS}$ without classification loss (Eqn.\ref{eq:sgan2}).
    \item GeoGAN$_{noL_{shape}}$- GeoGAN$_{WSS}$ without shape relationship  modeling term (Eqn.\ref{eq:sgan3}).
    \item GeoGAN$_{NoSamp}$ - GeoGAN$_{WSS}$ without  uncertainty sampling for injecting diversity to determine sampling's relevance to the final network performance. 
    %The original and upsampled images are directly connected without any sampling step.
    %
    %
    % \item  GeoGAN$_{L_{class}}$ - GeoGAN using classification loss (Eqn.\ref{eq:sgan2}) and adversarial loss (Eqn.\ref{eq:D2}) to determine $L_{class}$'s  relevance to GeoGAN's performance.
    % \item GeoGAN$_{L_{shape}}$ - GeoGAN using shape loss (Eqn.\ref{eq:sgan3}) and adversarial loss (Eqn.\ref{eq:D2}) to determine $L_{shape}$'s contribution to GeoGAN's performance.
    % \item GeoGAN$_{Samp}$ - GeoGAN using only uncertainty sampling and adversarial loss (Eqn.\ref{eq:D2}). This baseline quantifies the contribution of sampling to the image generation process. 
\end{enumerate}

\subsection{Efficacy of Weakly Supervised Semantic Segmentation}

\begin{figure}[h]
\centering
\begin{tabular}{ccc}
\includegraphics[height=2.4cm, width=2.6cm]{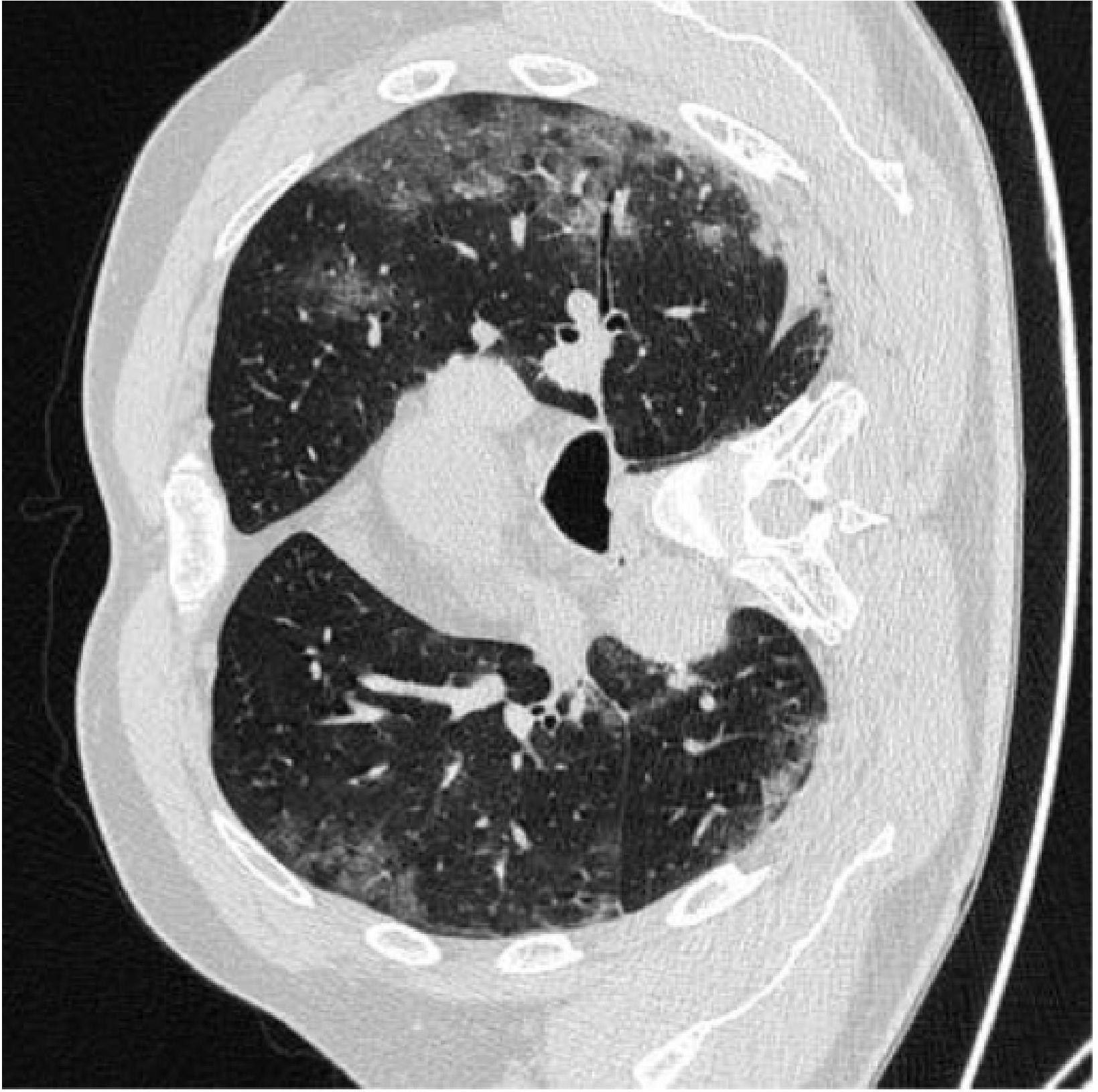} & 
\includegraphics[height=2.4cm, width=2.6cm]{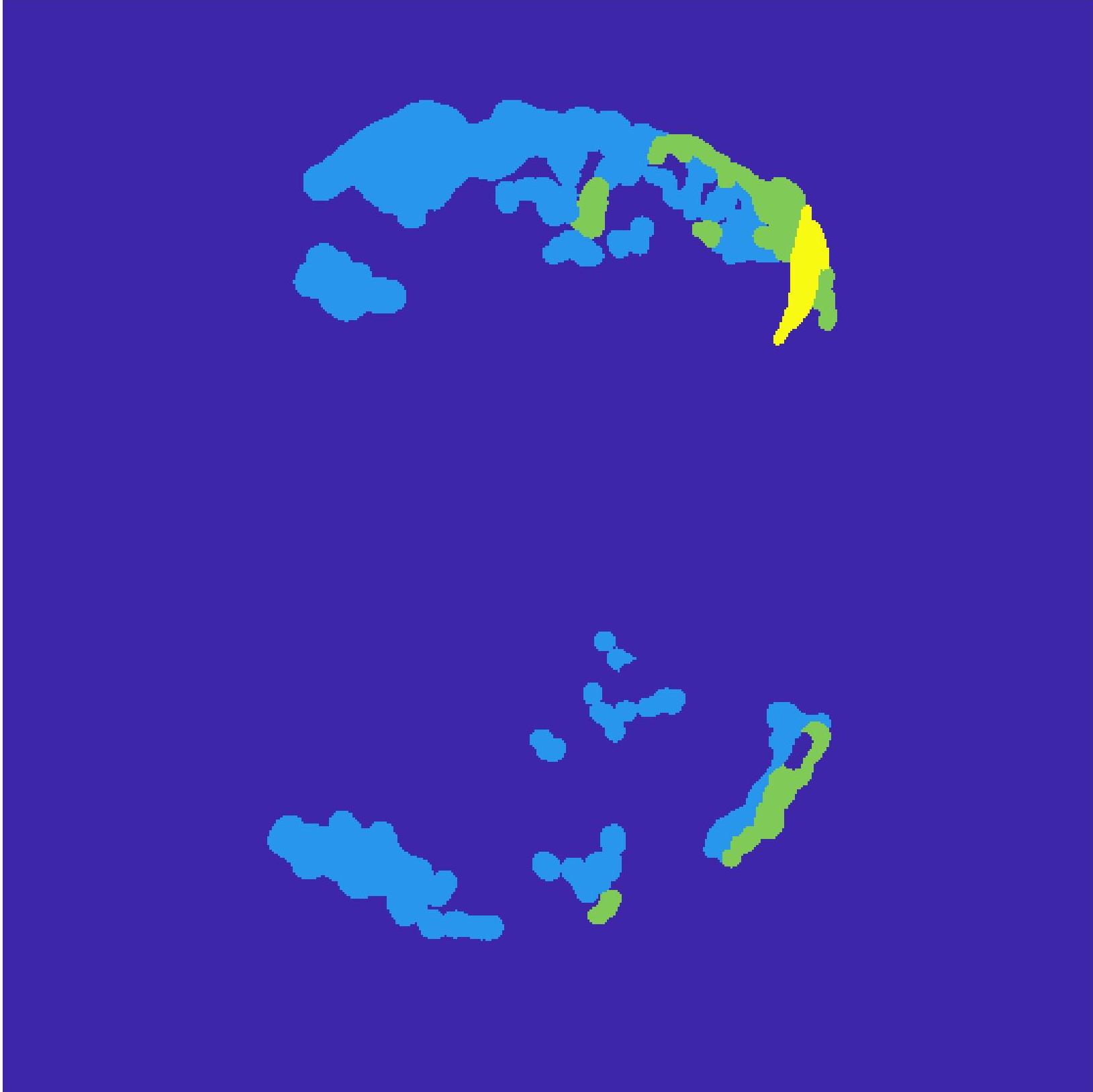} & 
\includegraphics[height=2.4cm, width=2.6cm]{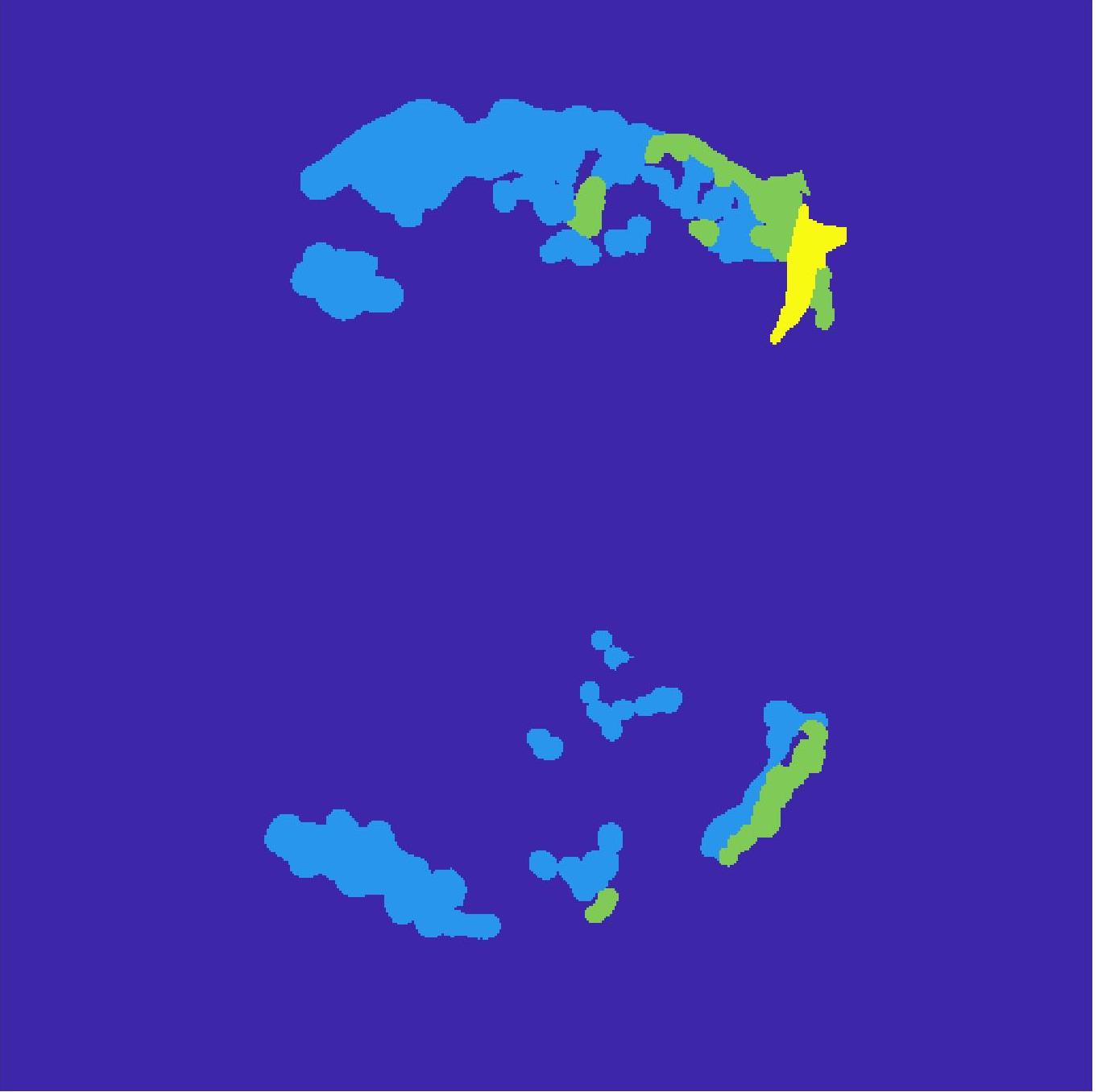} \\
%---------
\includegraphics[height=2.4cm, width=2.6cm]{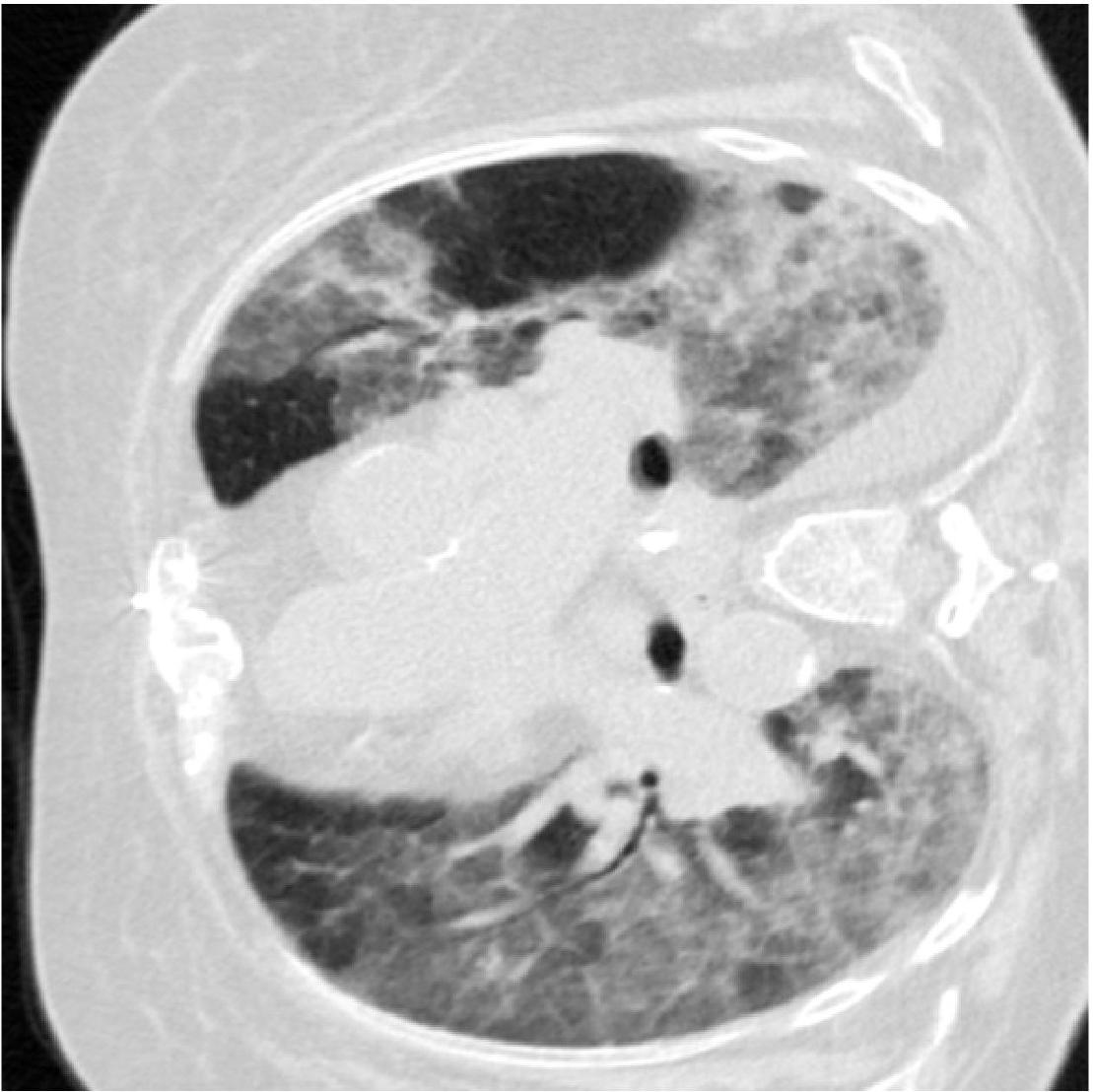} & 
\includegraphics[height=2.4cm, width=2.6cm]{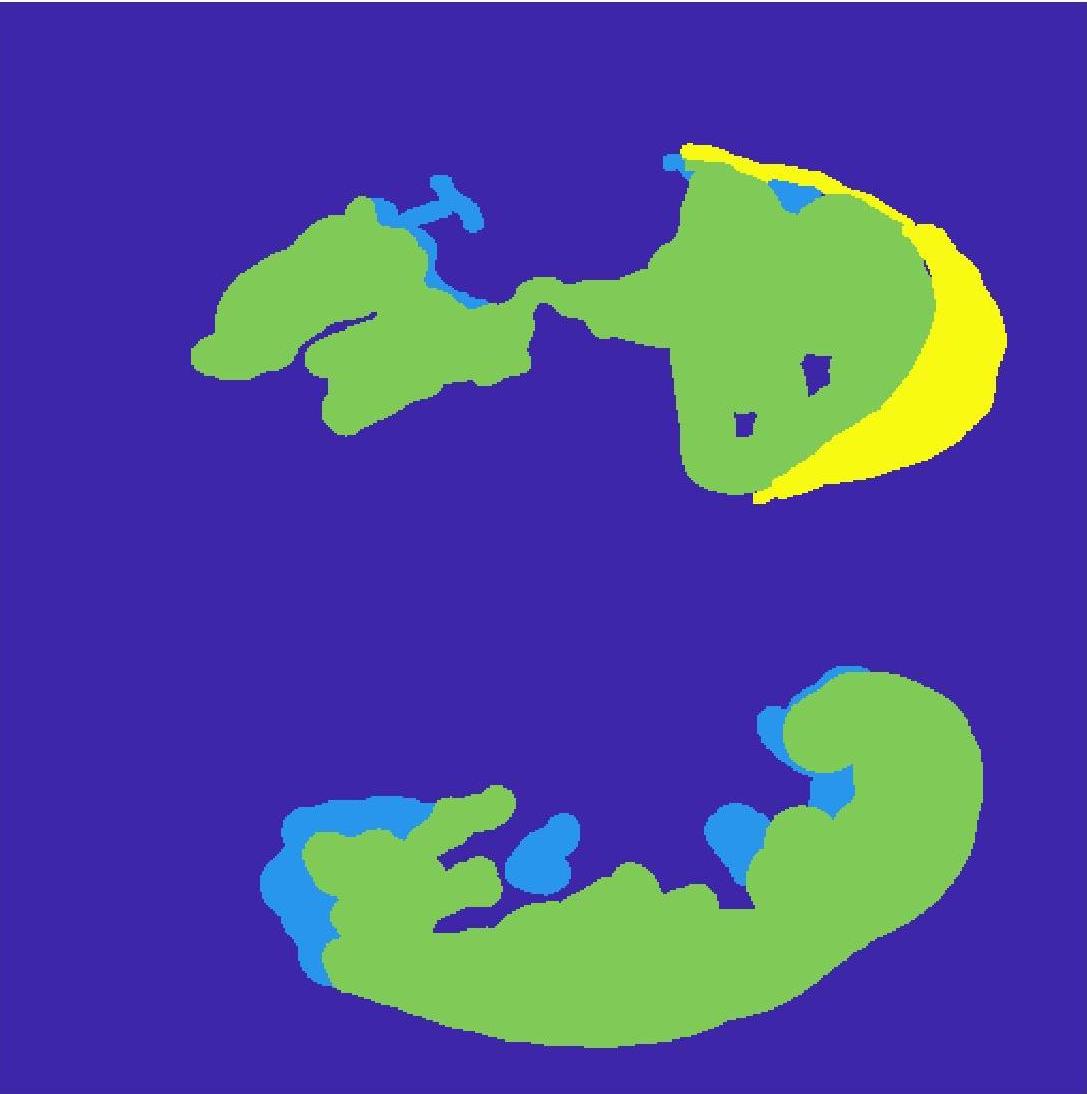} & 
\includegraphics[height=2.4cm, width=2.6cm]{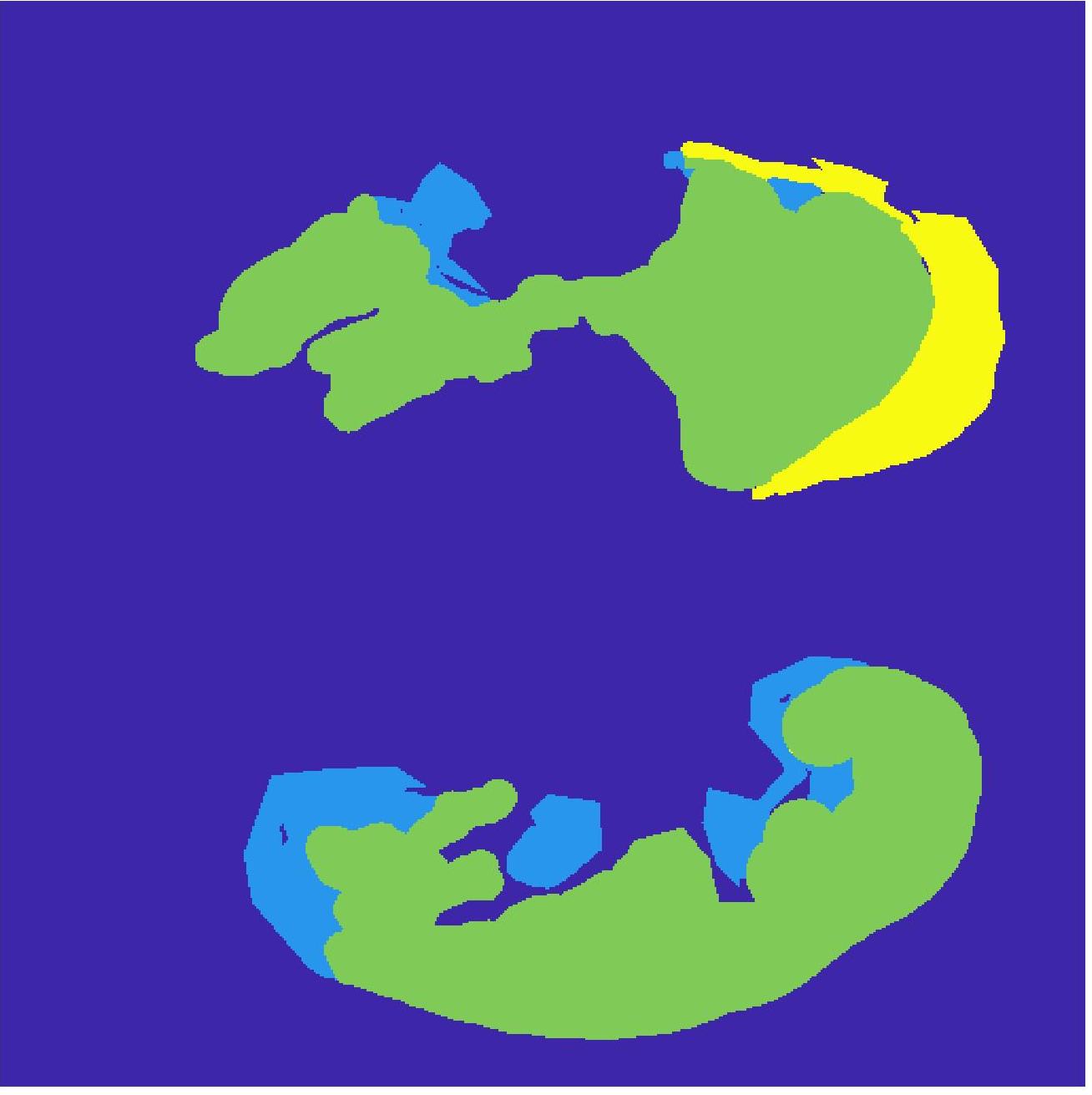} \\
(a) & (b) & (c) \\
\end{tabular}
\caption{Results for weakly supervised segmentation compared to ground truth maps. (a) CT image; (b) manual ground truth segmentation label map; (c) label map generated by our weakly supervised approach. Light blue is label $1$ - ground glass opacity; Green is label $2$ - consolidation; Yellow is label $3$- pleural effusion.}
\label{fig:WSS}
\end{figure}

To quantify the accuracy of the weakly supervised segmentation step we compare the segmentation output with the manual segmentations and obtain $DM=0.926$, and $HD=6.5$ mm. These numbers indicate very good agreement with the expert's manual annotations. Figure~\ref{fig:WSS} shows two examples of an image, its ground truth label map, and the label map obtained from the weakly supervised segmentation step. We observe that the WSS map closely resembles the manual map with small regions of oversegmentation. 
Oversegmented label maps recover the entire annotated area and hence enable GeoGAN to learn the full range of geometrical relations between different labels. On the other hand, oversegmented maps also let the algorithm learn noisy features due to the fact that non-diseased regions are included as infections. %In the case of undersegmented maps GeoGAN may not learn the whole set of features modeling the relation between different labeled regions.

\subsection{Segmentation Results And Analysis}
\label{expt:seg}

We hypothesize that a good image augmentation method should capture the different complex relationships between different labels, with the generated images leading to improvement in segmentation accuracy. 
Table~\ref{tab:CTSeg1} shows the average DM and MAE for different augmentation methods on the $CTSeg1$ dataset for infection regions. We also report the performance of the recent approach of \cite{InfNet} on the same dataset.

Table~\ref{tab:CTSeg1} shows the $p$ values comparing the results of all methods with $GeoGAN_{WSS}$ (except \cite{InfNet} since we do not have access to all segmentation masks). Results of $GeoGAN_{Manual}$ denote the best performance obtained with a given network since they are trained on clinician provide manual segmentation maps. $GeoGAN_{WSS}$'s results show that the WSS component is very accurate in obtaining semantic segmentation and can be used effectively where manual segmentation maps are unavailable. 

\begin{figure*}[h]
\centering
\begin{tabular}{cccccc}
\includegraphics[height=2.5cm, width=2.5cm]{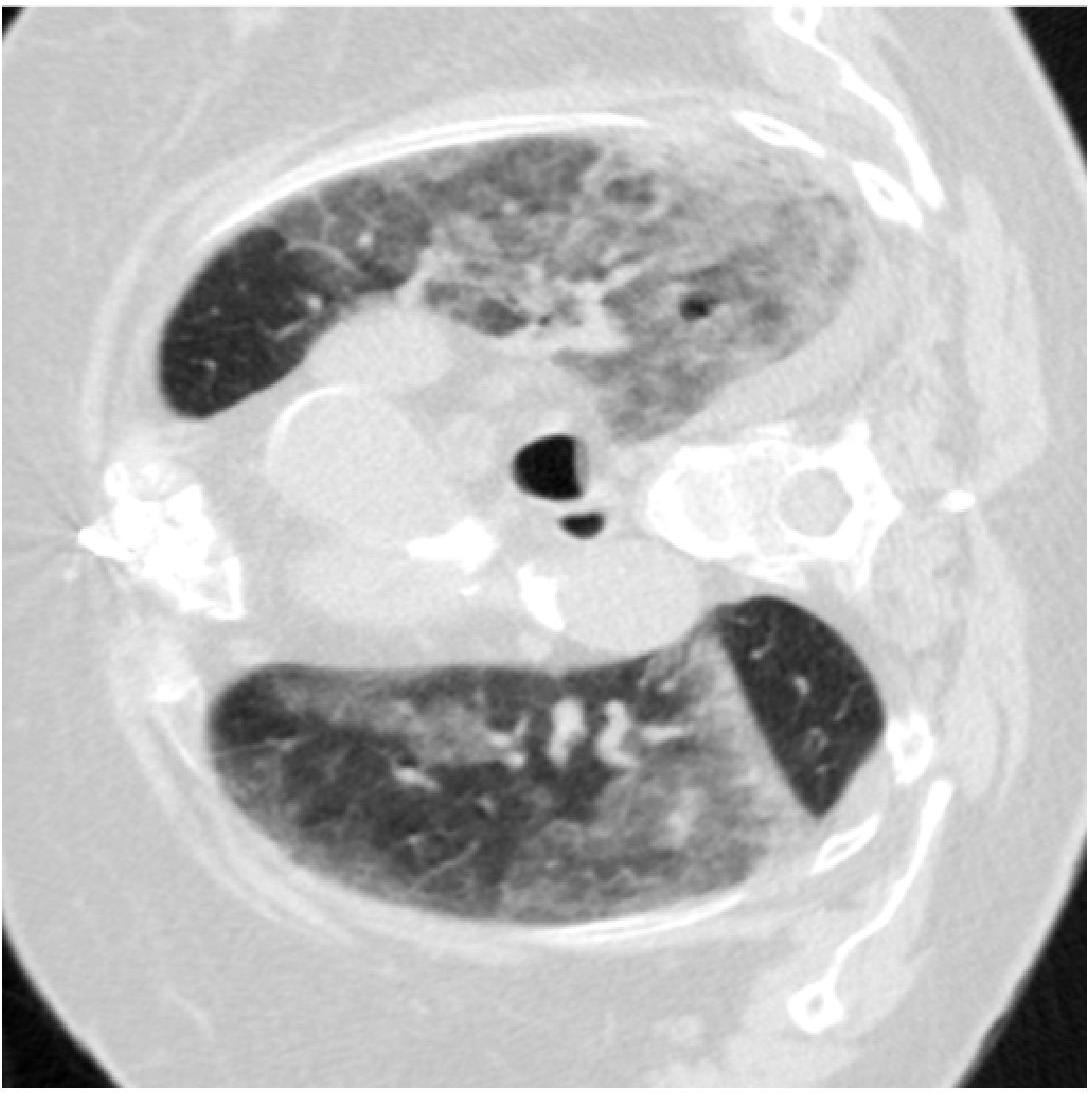} & 
\includegraphics[height=2.5cm, width=2.5cm]{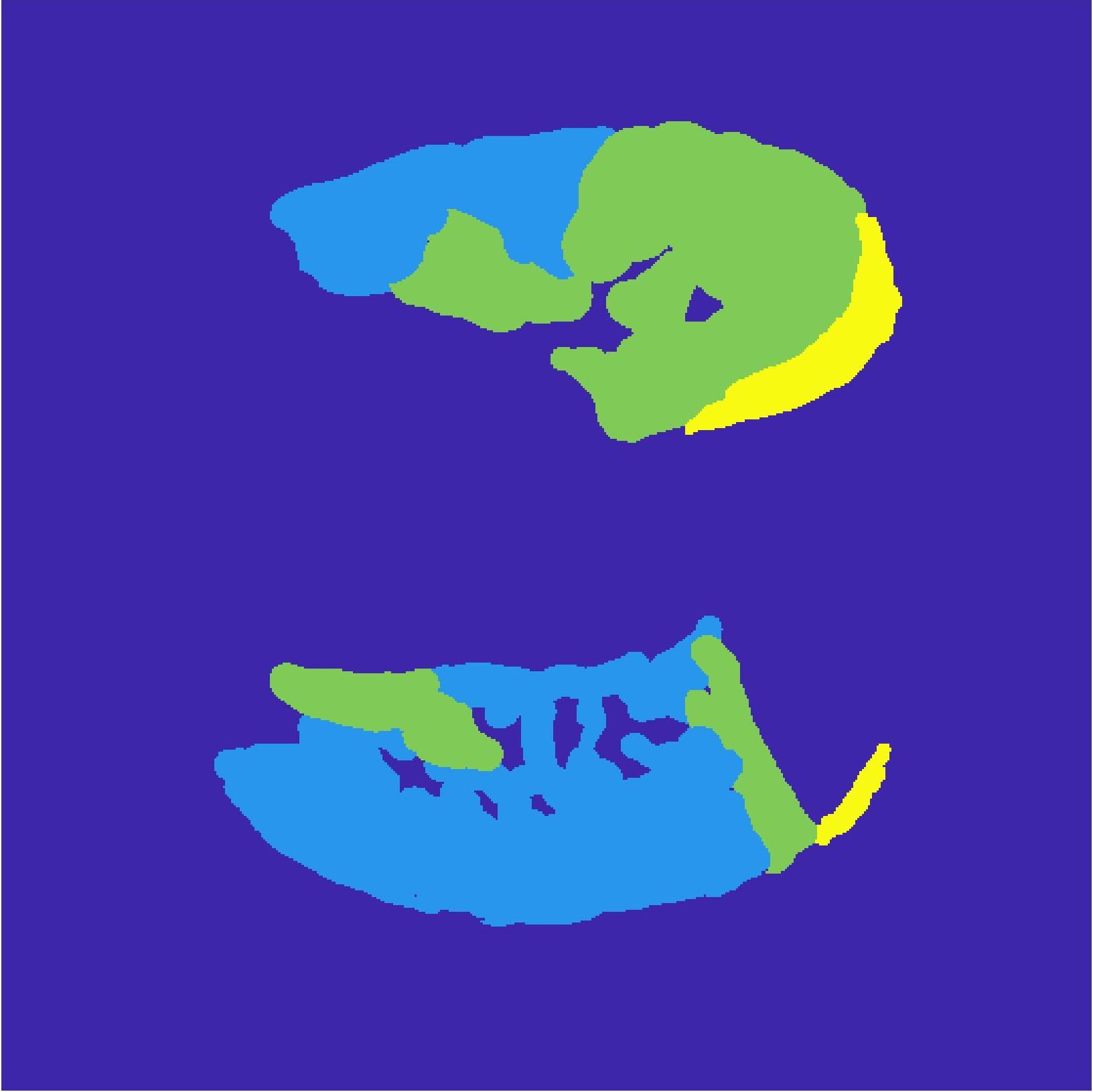} & 
\includegraphics[height=2.5cm, width=2.5cm]{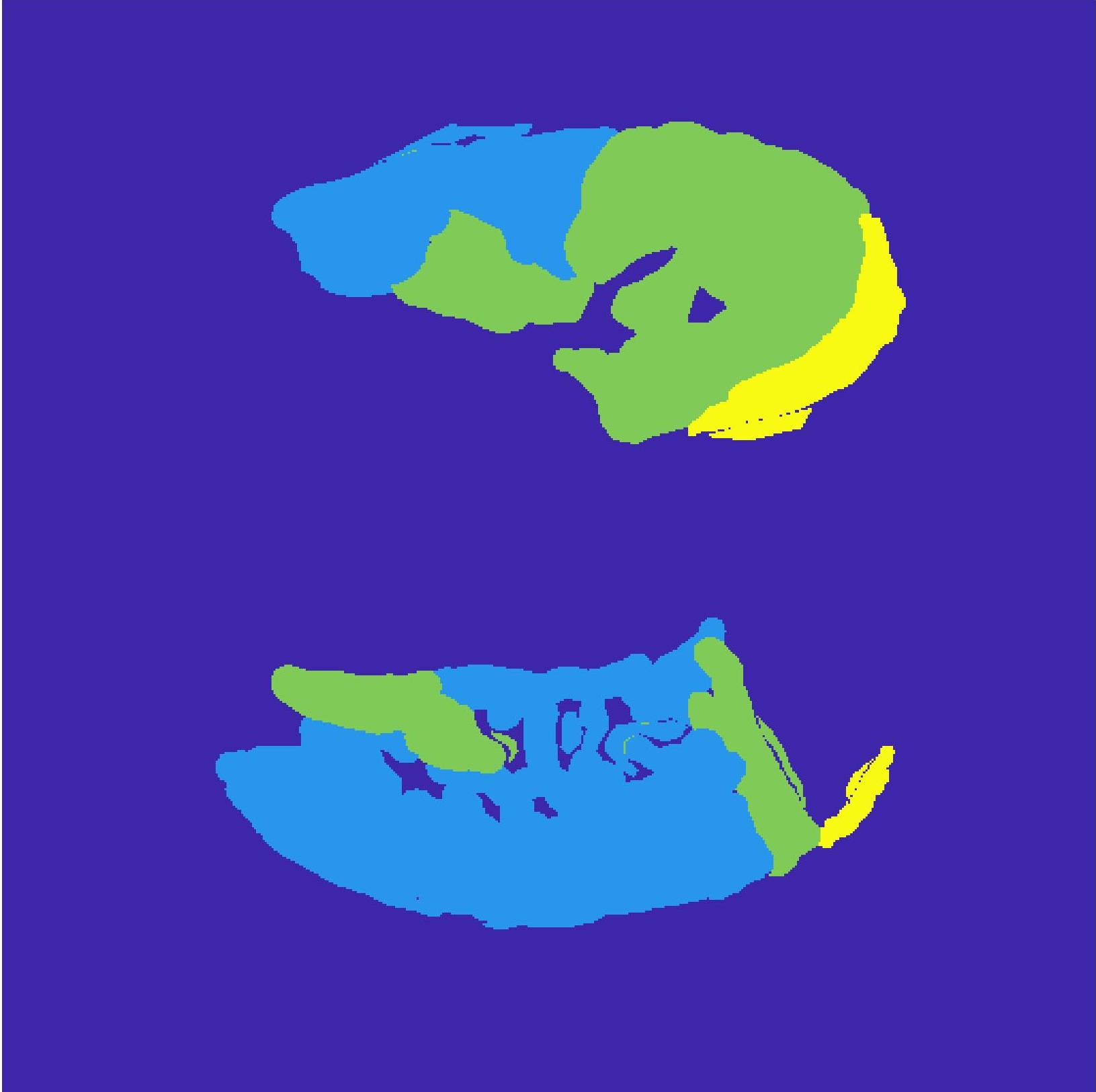} & 
\includegraphics[height=2.5cm, width=2.5cm]{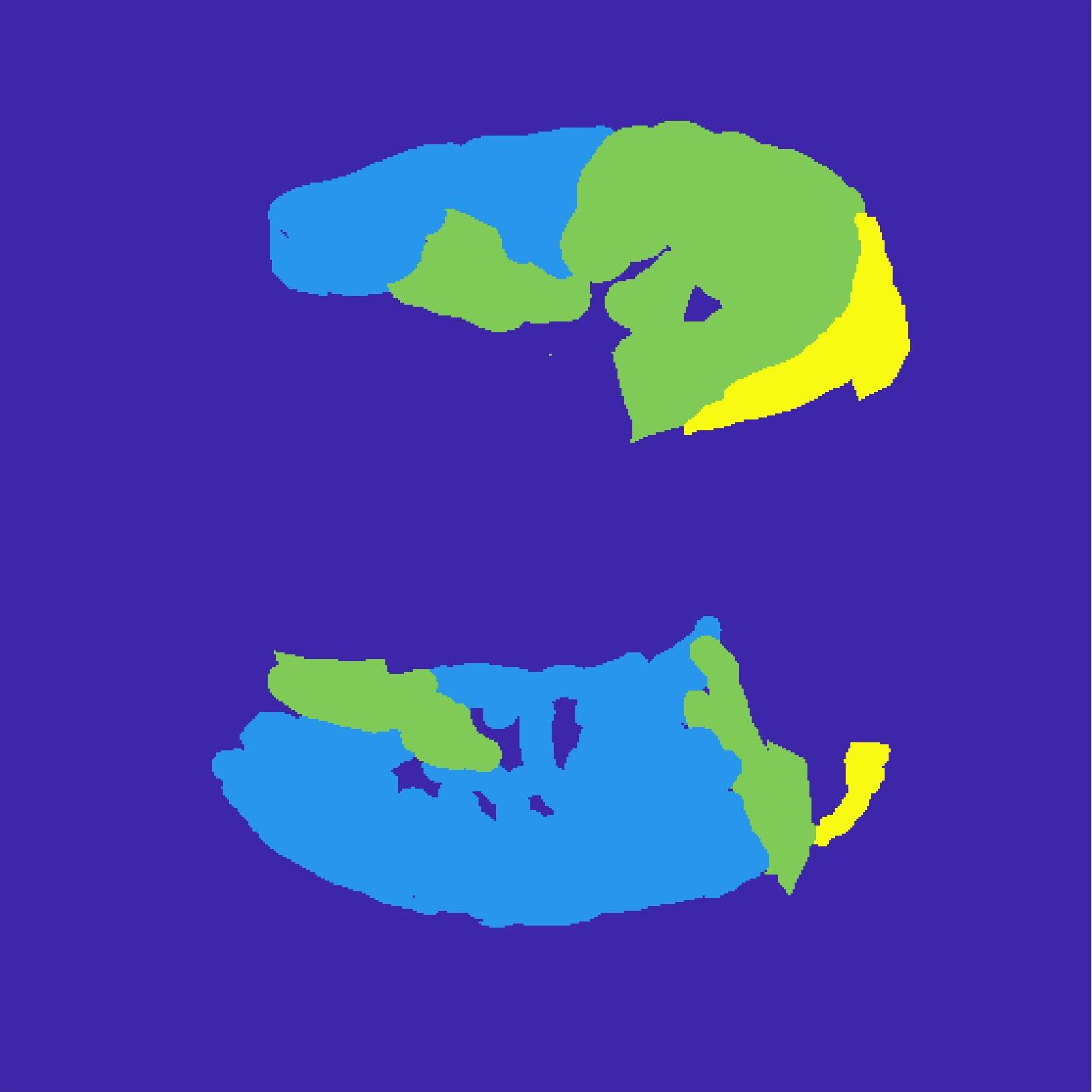} & 
\includegraphics[height=2.5cm, width=2.5cm]{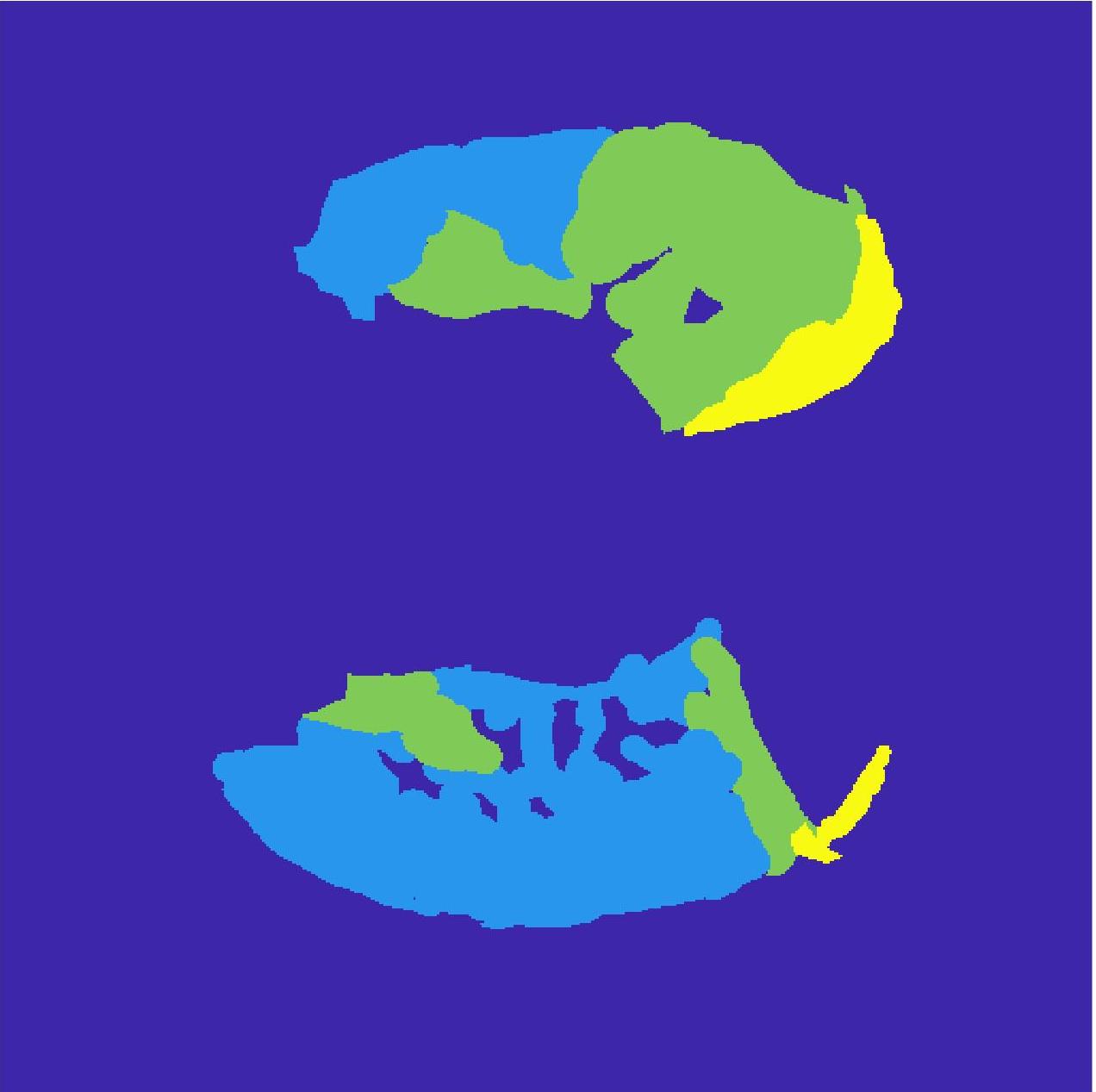} & 
\includegraphics[height=2.5cm, width=2.5cm]{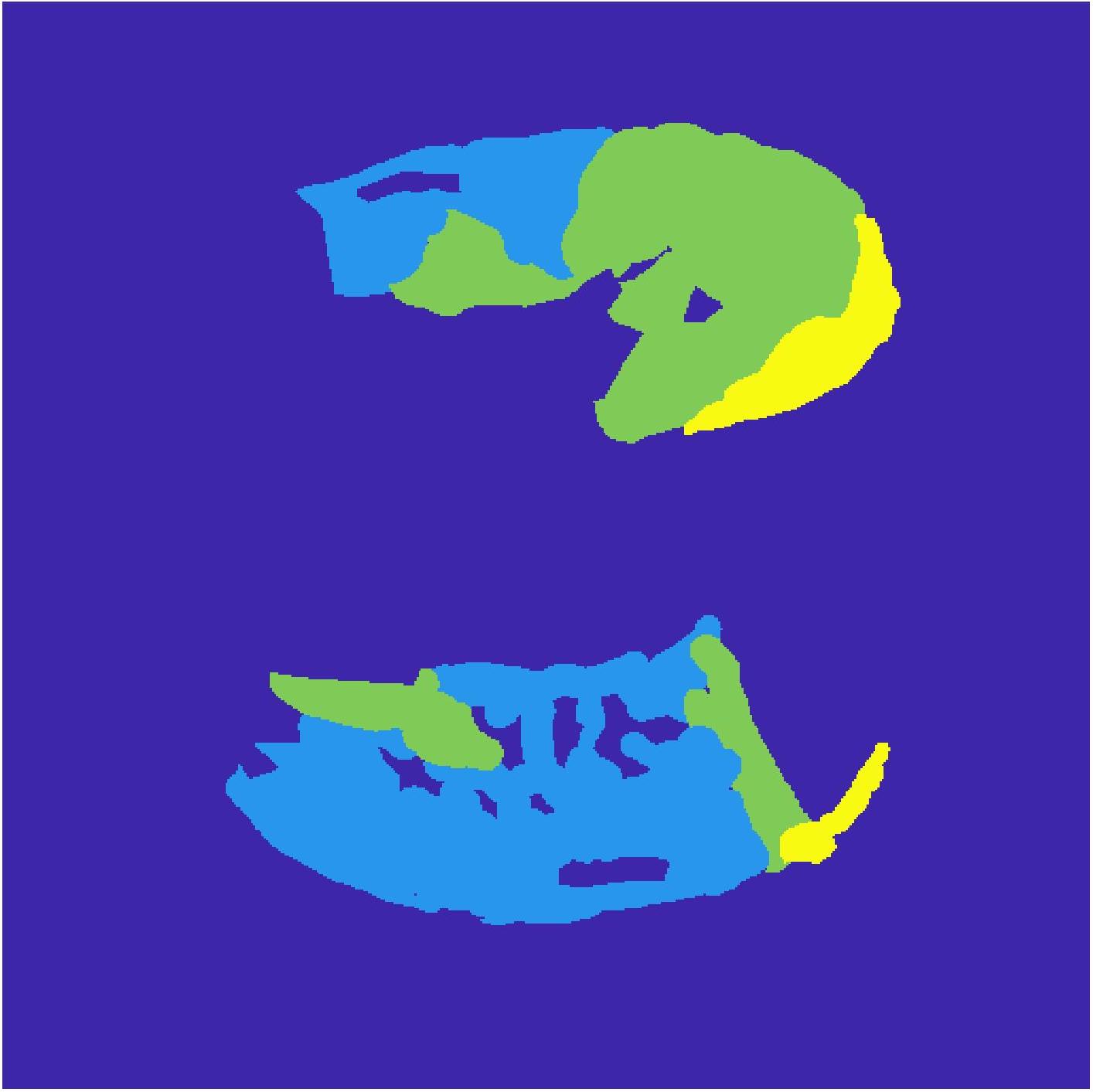} \\
%---------
\includegraphics[height=2.5cm, width=2.5cm]{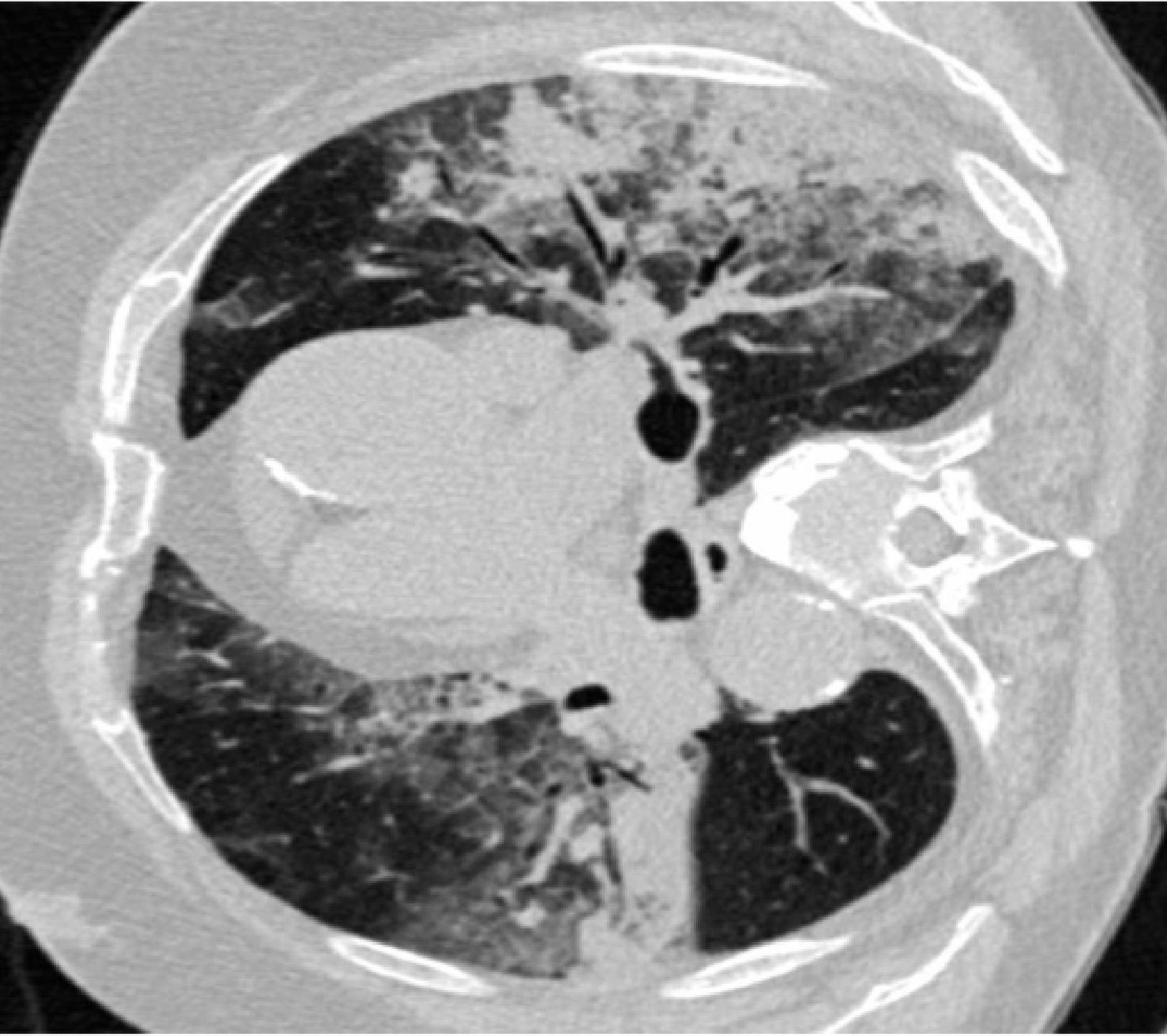} & 
\includegraphics[height=2.5cm, width=2.5cm]{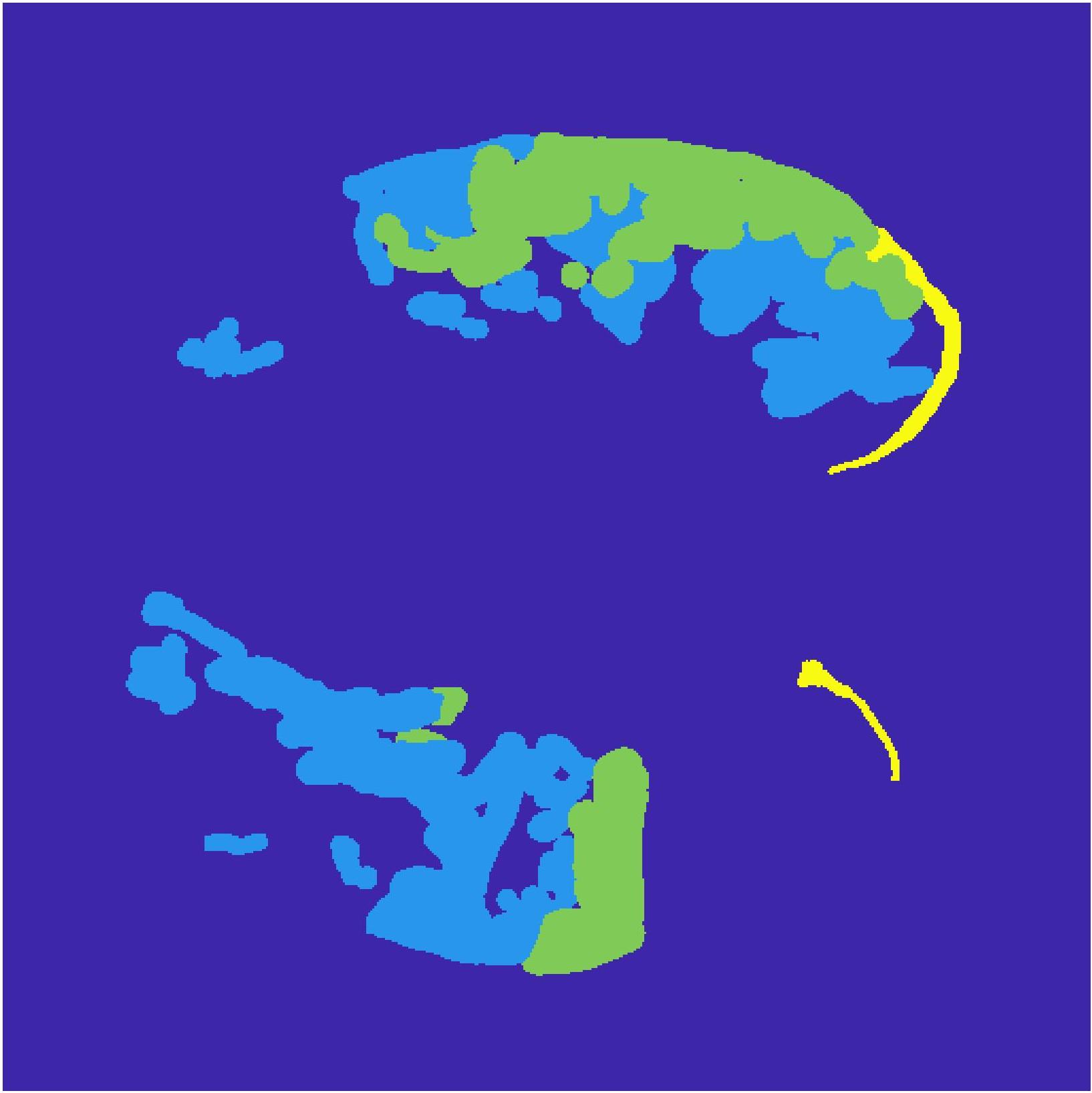} & 
\includegraphics[height=2.5cm, width=2.5cm]{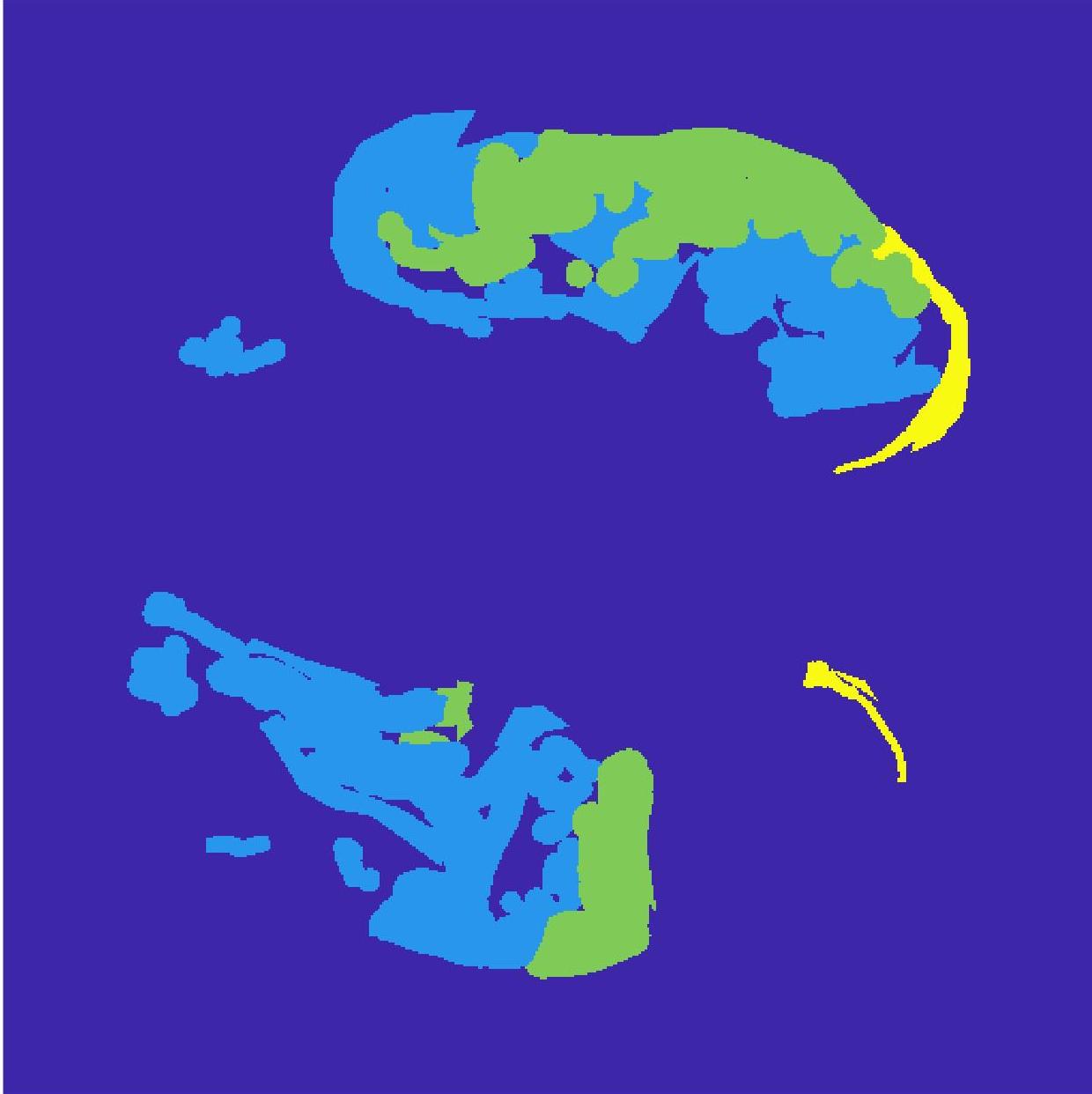} & 
\includegraphics[height=2.5cm, width=2.5cm]{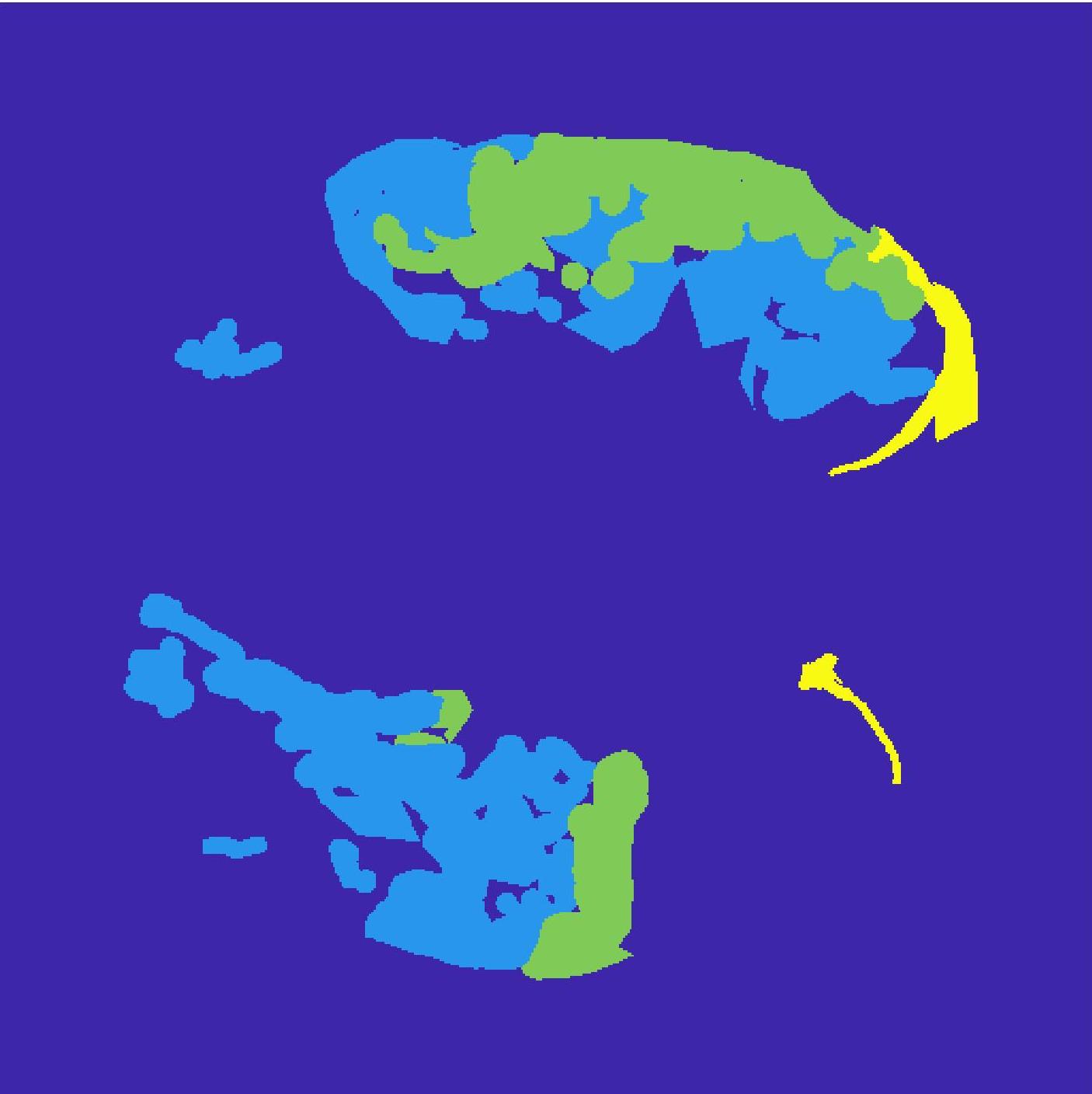} & 
\includegraphics[height=2.5cm, width=2.5cm]{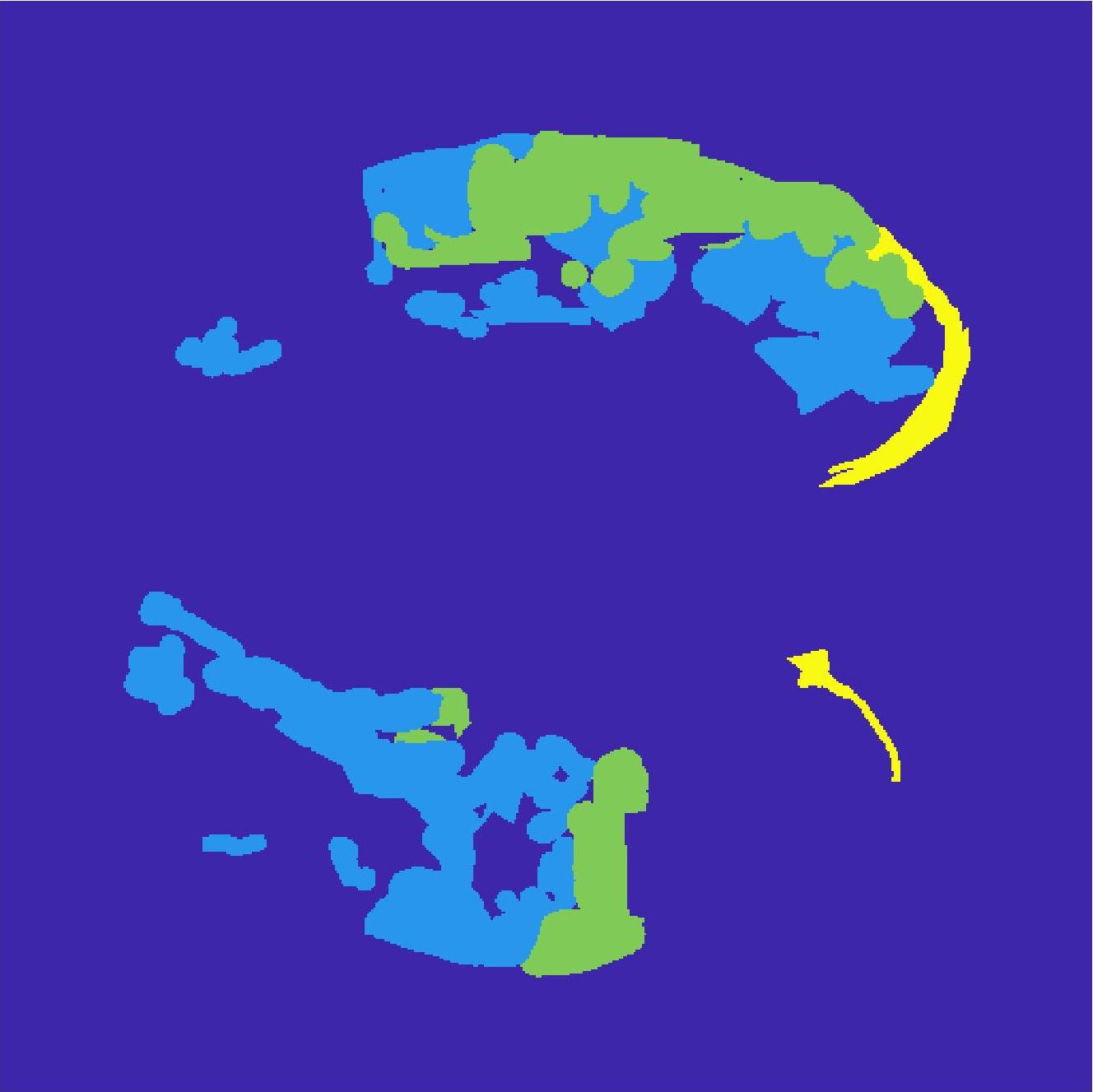} & 
\includegraphics[height=2.5cm, width=2.5cm]{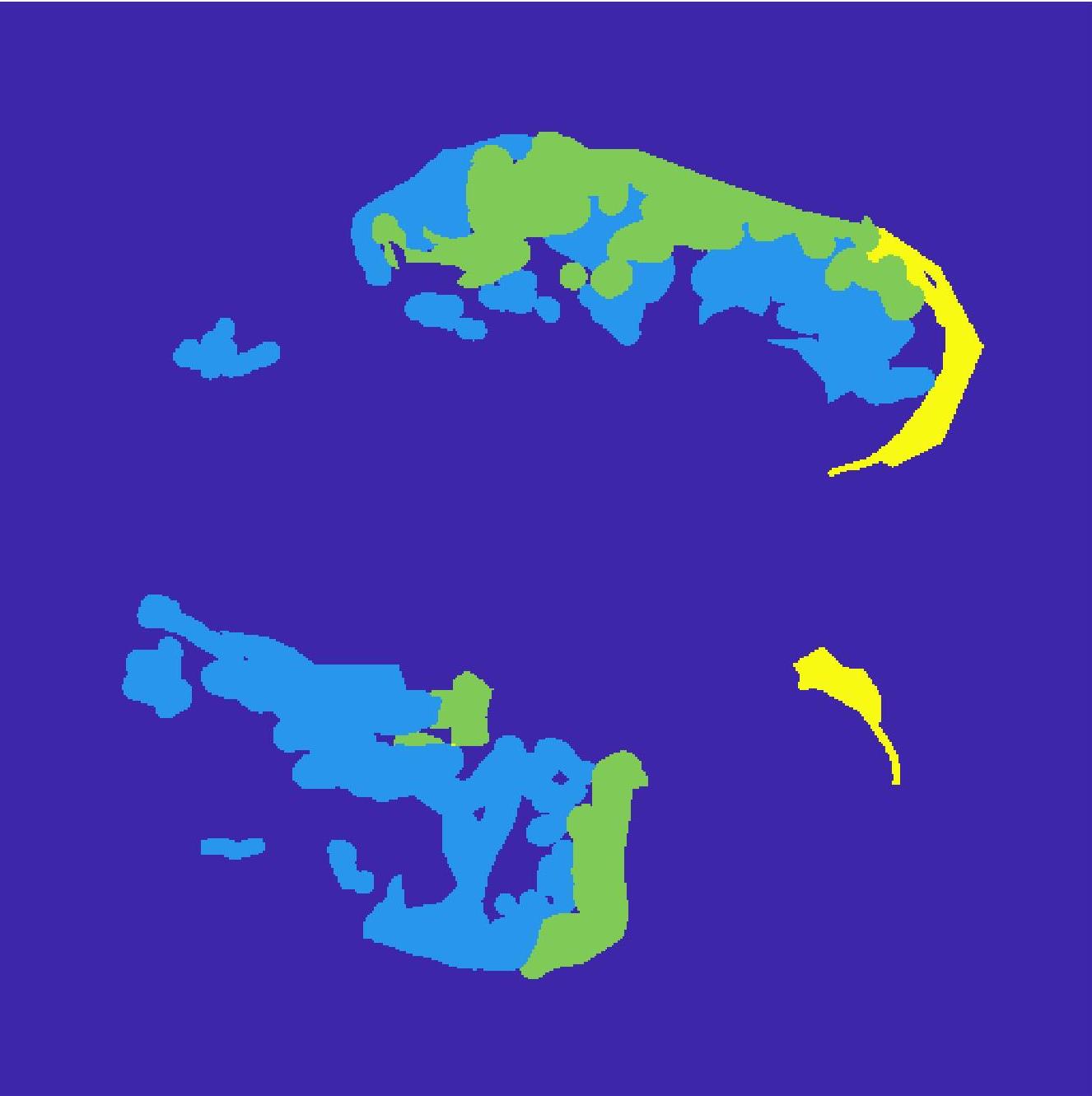} \\
(a) & (b) & (c) & (d) & (e) & (f) \\
\end{tabular}
\caption{Segmentation results on the $CTSeg1$ dataset: (a) original test  images; (b) manual segmentation masks. Masks generated using data generated by: (c) GeoGAN; (d) Zhao \cite{Zhao_CVPR2019}; (e) $DAGAN$; (f) $cGAN$. The two rows correspond to two different images.}
\label{fig:segout1}
\end{figure*}

Figure~\ref{fig:segout1} shows the segmentation results using a UNet++ trained on images from different methods. Figure~\ref{fig:segout1} (a) shows the test image and Figure~\ref{fig:segout1} (b) shows the manual mask. Figures~\ref{fig:segout1} (c)-(f) show, respectively, the segmentation masks obtained by GeoGAN$_{WSS}$, \cite{Zhao_CVPR2019}, $DAGAN$ and $cGAN$.

Subsequently, in Tables~\ref{tab:CTSeg2},\ref{tab:CTSeg3} we show results of $GeoGAN_{WSS}$ (denoted as $GeoGAN$ for brevity) for datasets $CTSeg2$ and $CTSeg3$. 
Our method outperforms baseline conventional data augmentation and other competing methods by a significant margin. GeoGAN$_{WSS}$'s DM is higher than the DM value of the best performing method.  We also perform better than \cite{InfNet} on $CTSeg1$, which uses semi-supervised approaches to outperform a UNet++ architecture (using conventional data augmentation). Our results clearly show that with better augmentation techniques we can do better than state of the art segmentation network architectures.

GeoGAN's superior segmentation accuracy is attributed to it's capacity to learn geometrical relationship between different labels (through $L_{shape}$) much better than competing methods. Thus our attempt to model the intrinsic geometrical relationships between different labels could generate superior quality masks. %

\begin{table*}[h]
 \begin{center}
\begin{tabular}{|c|c|c|c|c|c|c|c|}
\hline 
& \multicolumn{4}{|c|}{Comparison approaches} & \multicolumn{2}{|c|}{Proposed} & {}\\ \hline
{} & {DA}  & {DAGAN}  & {cGAN} & {Zhao} & {GeoGAN$_{WSS}$} & {GeoGAN$_{Manual}$} &{\cite{InfNet}} \\ \hline
%{}& {}  & {\cite{DAGAN}} & {\cite{Mahapatra_MICCAI2018}} & {\cite{Zhao_CVPR2019}} & {} & {} \\ \hline
%
{DM} & {0.704} & {0.719} & {0.738} & {0.752} & {0.781} & {\textbf{0.789}} & {0.764} \\ 
{} & {(0.17)} & {(0.12)} & {(0.09)} & {(0.08)} & {(0.05)} & {(\textbf{0.03})} & {-}\\ \hline
{p} & {0.006} & {0.004}  & {0.0005} & {0.0003} & {-} & {0.11} & {-}\\ \hline
{MAE} & {0.097} & {0.088}  & {0.083} & {0.071} & {0.058} & {\textbf{0.053}} & {0.064}\\ 
{} & {(0.018)} & {(0.015)}  & {(0.015)} & {(.017)} & {(.013)} & {(\textbf{.012})} & {-} \\ \hline
{HD} & {13.9} & {12.7}  & {10.9} & {9.1} & {8.3} & {\textbf{8.2}} & {-}\\ 
{} & {(4.1)} & {(3.9)}  & {(3.8)} & {(3.1)} & {(2.4)} & {(\textbf{2.3})} &{-}\\ \hline
\end{tabular}
\caption{COVID Segmentation results for $CTSeg1$ dataset. Mean and standard deviation (in brackets) are shown. Best results per metric is shown in bold. $p$ values are with respect to $GeoGAN_{WSS}$}
\label{tab:CTSeg1}
\end{center}
\end{table*}

\begin{table}[h]
 \begin{center}
\begin{tabular}{|c|c|c|c|c|c|}
\hline 
& \multicolumn{4}{|c|}{Comparison approaches} & Proposed \\ \hline
{} & {DA}  & {DAGAN}  & {cGAN} & {Zhao} & {GeoGAN} \\ 
% {}& {}  & {\cite{DAGAN}} & {\cite{Mahapatra_MICCAI2018}} & {\cite{Zhao_CVPR2019}} & {}  \\ 
\hline
{DM} & {0.708} & {0.727} & {0.761} & {0.774} & {\textbf{0.809}} \\ 
{} & {(0.15)} & {(0.13)} & {(0.10)} & {(0.09)} & {(\textbf{0.07})} \\ \hline
{HD} & {14.2} & {12.4}  & {11.2} & {9.4} & {\textbf{8.7}} \\ 
{} & {(4.4)} & {(3.7)}  & {(3.2)} & {(3.0)} & {(\textbf{2.7})} \\ \hline
{MAE} & {0.112} & {0.094}  & {0.088} & {0.079} & {\textbf{0.073}} \\
{} & {(0.011)} & {(0.009)}  & {(0.010)} & {(0.008)} & {\textbf{(0.006)}} \\\hline
{p} & {0.0007} & {0.005}  & {0.0001} & {0.01} & {-} \\ \hline
\end{tabular}
\caption{COVID Segmentation results for $CTSeg2$ dataset. Mean and standard deviation (in brackets) are shown. Best results per metric is shown in bold. $p-$values are with respect to GeoGAN.}% HD is in mm}
\label{tab:CTSeg2}
\end{center}
\end{table}

\begin{table}[h]
 \begin{center}
\begin{tabular}{|c|c|c|c|c|c|}
\hline 
& \multicolumn{4}{|c|}{Comparison approaches} & Proposed \\ \hline
{} & {DA}  & {DAGAN}  & {cGAN} & {Zhao} & {GeoGAN} \\ 
% {}& {}  & {\cite{DAGAN}} & {\cite{Mahapatra_MICCAI2018}} & {\cite{Zhao_CVPR2019}} & {}  \\ 
\hline
{DM} & {0.723} & {0.762} & {0.779} & {0.793} & {\textbf{0.815}} \\ 
{} & {(0.12)} & {(0.10)} & {(0.13)} & {(0.08)} & {(\textbf{0.05})} \\ \hline
{HD} & {13.7} & {12.4}  & {11.2} & {9.0} & {\textbf{7.8}} \\ 
{} & {(4.2)} & {(3.6)}  & {(3.9)} & {(3.3)} & {(\textbf{3.1})} \\ \hline
{MAE} & {0.097} & {0.089}  & {0.081} & {0.077} & {\textbf{0.069}} \\  {} & {(0.009)} & {(0.010)}  & {(0.008)} & {(0.006)} & {\textbf{(0.005)}} \\\hline
{p} & {0.008} & {0.003}  & {0.009} & {0.01} & {-} \\ \hline
\end{tabular}
\caption{COVID Segmentation results for $CTSeg3$ dataset. Mean and standard deviation (in brackets) are shown. Best results per metric is shown in bold.$p-$values are with respect to GeoGAN.}% HD is in mm}
\label{tab:CTSeg3}
\end{center}
\end{table}

\subsection{Ablation Studies.}
\begin{figure}[h]
\centering
\begin{tabular}{ccc}
\includegraphics[height=2.5cm, width=2.5cm]{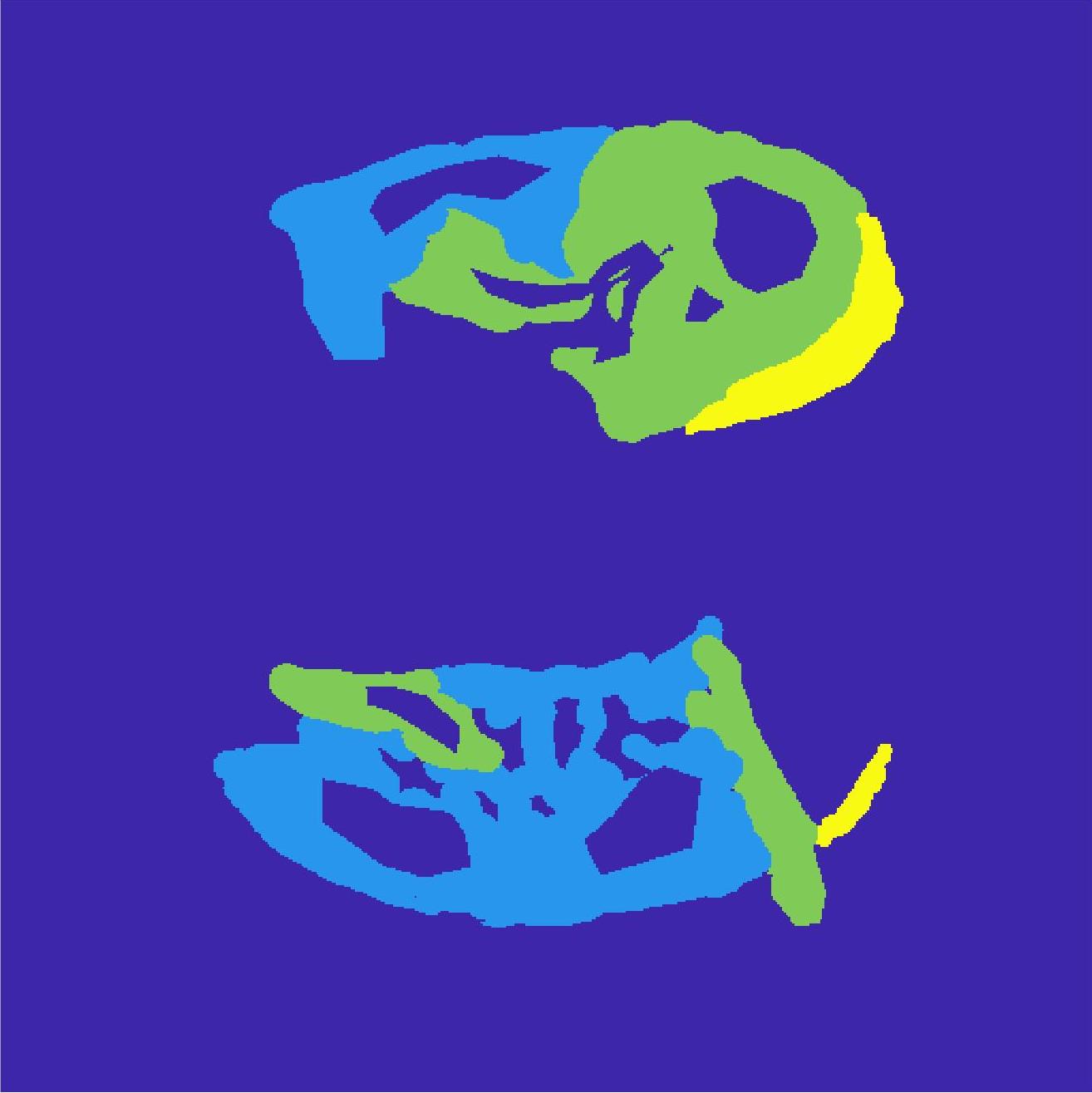} & 
\includegraphics[height=2.5cm, width=2.5cm]{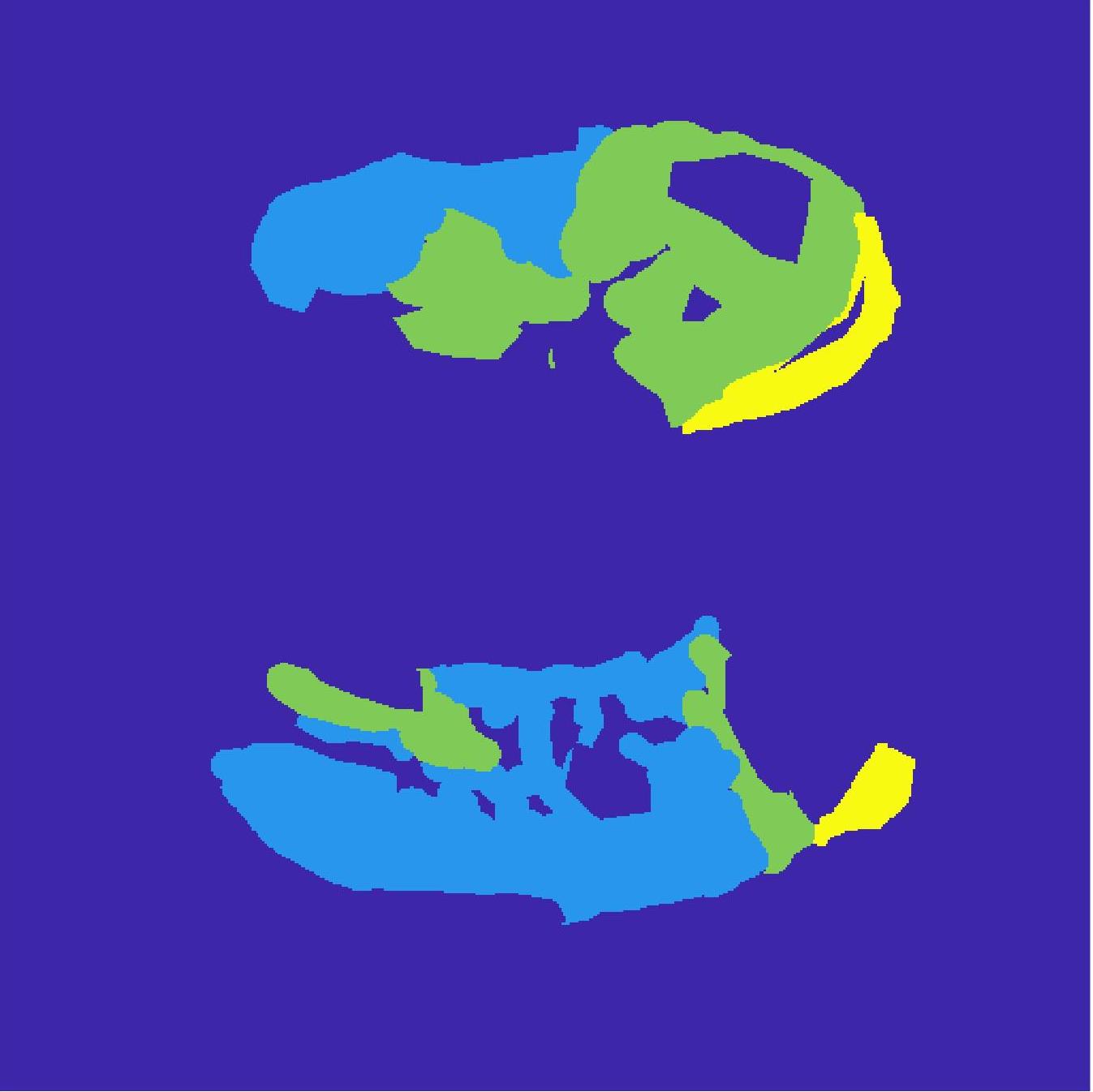} & 
\includegraphics[height=2.5cm, width=2.5cm]{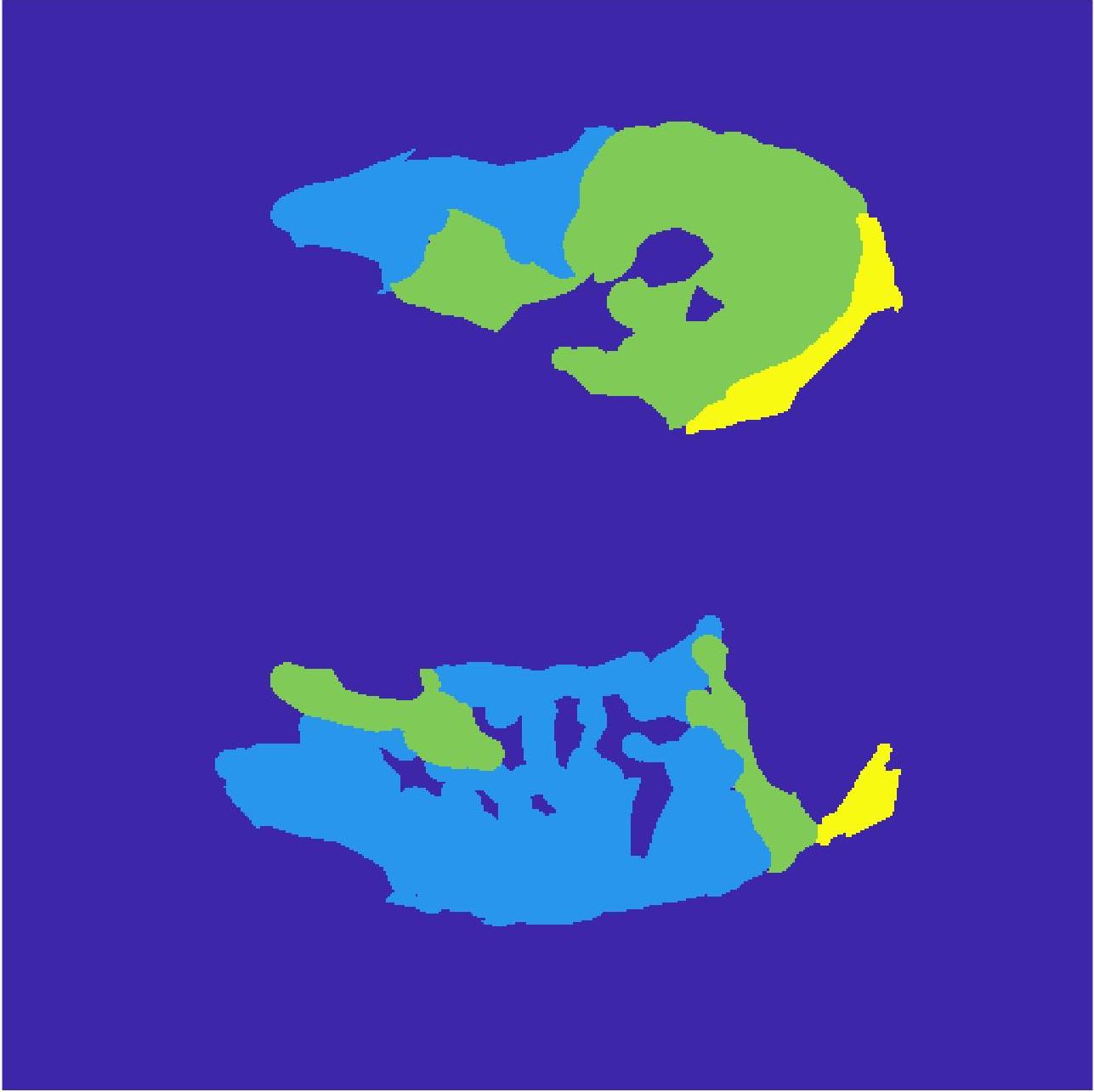} \\
%--------
\includegraphics[height=2.5cm, width=2.5cm]{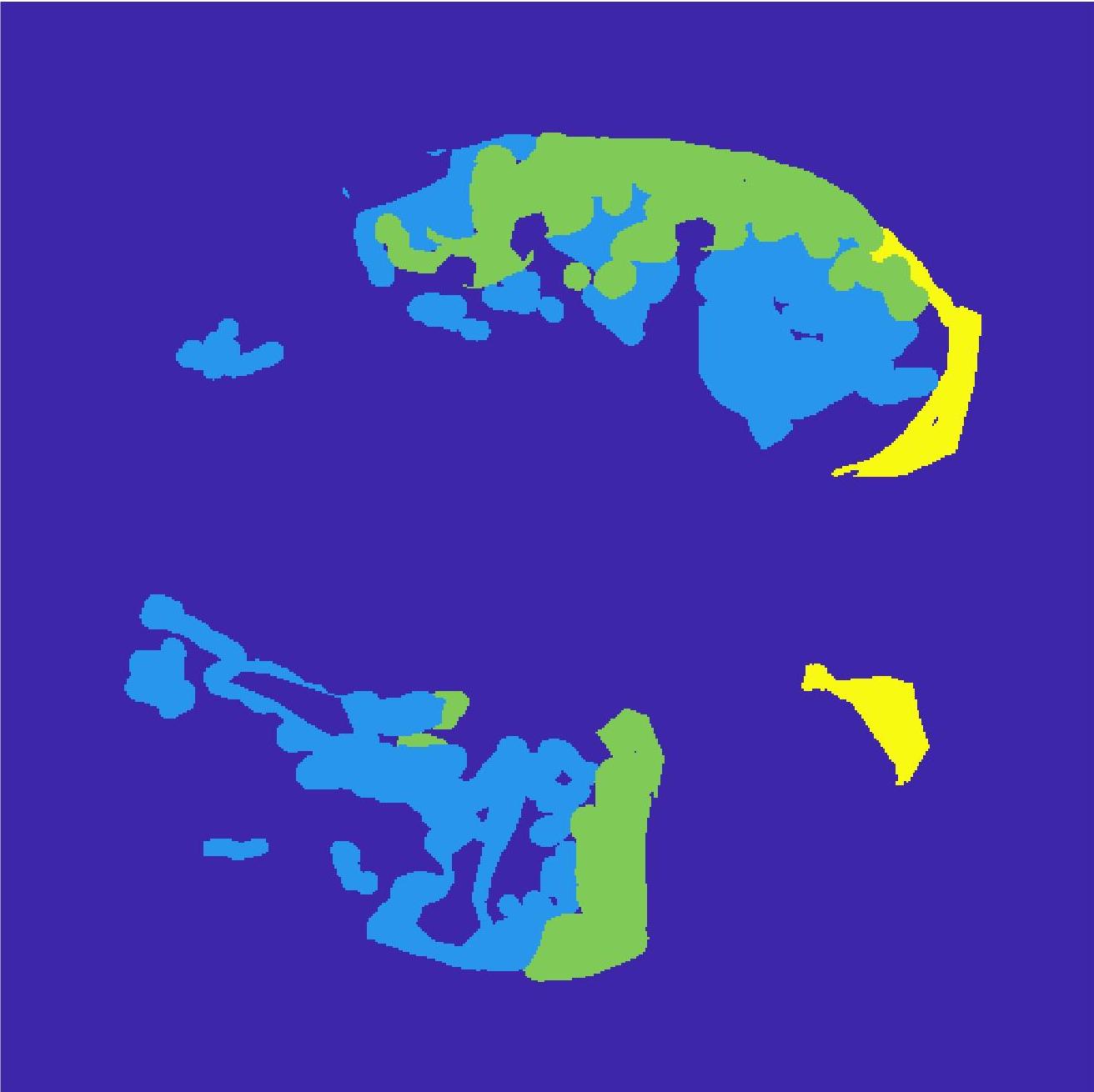} & 
\includegraphics[height=2.5cm, width=2.5cm]{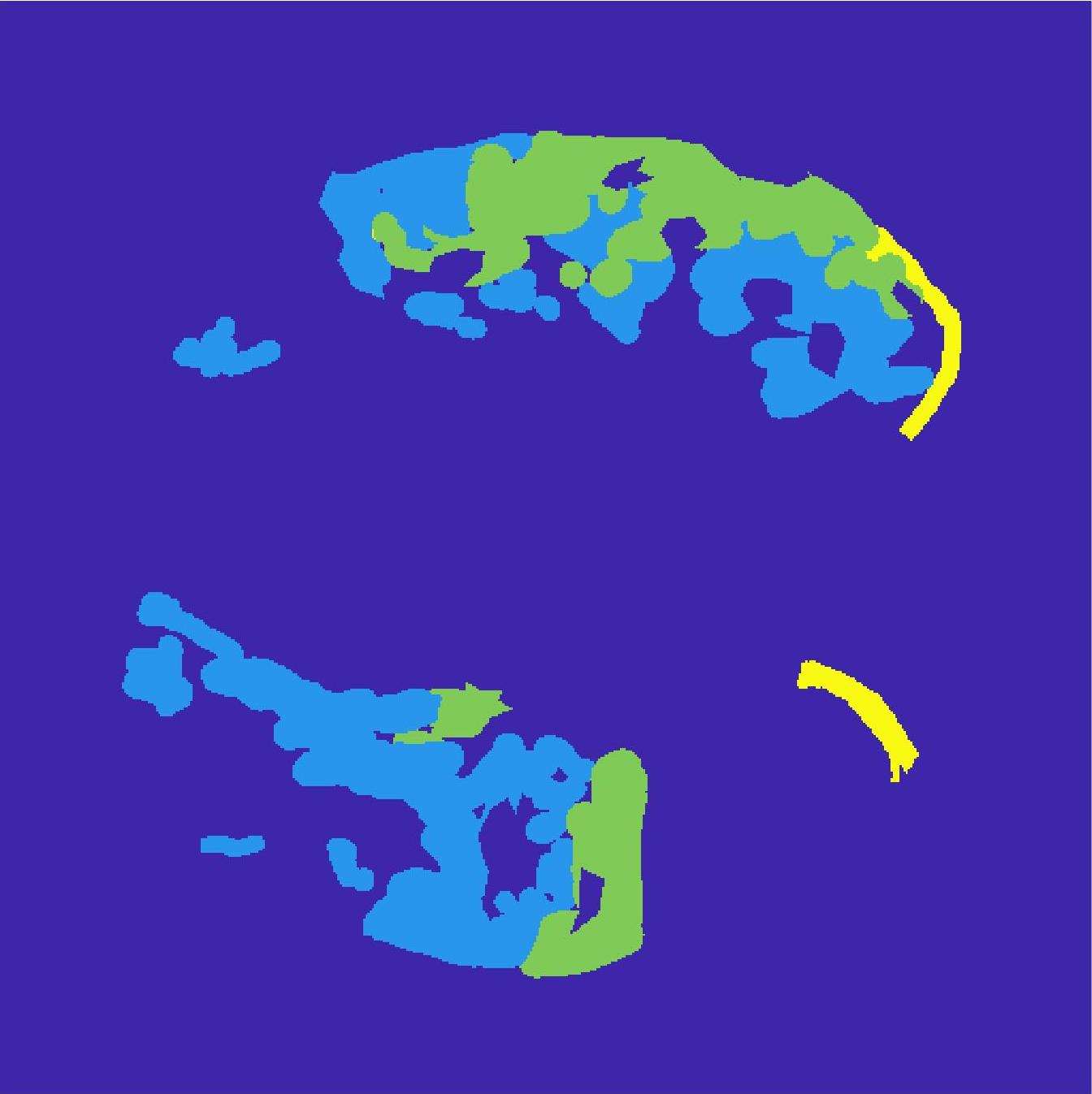} & 
\includegraphics[height=2.5cm, width=2.5cm]{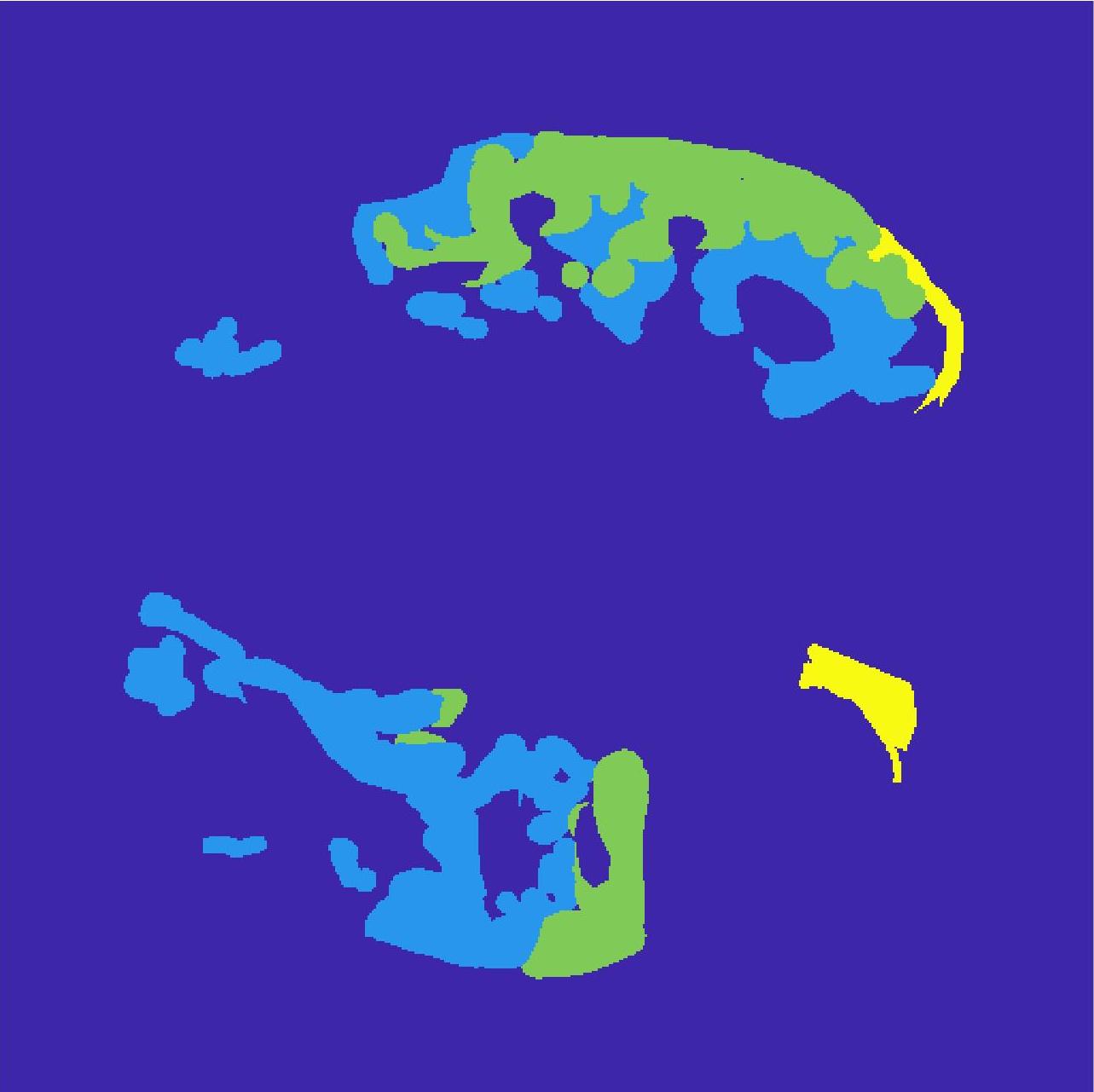} \\
 (a) & (b) & (c) \\
\end{tabular}
\caption{Segmentation results for ablation experiments on $CTSeg1$: (a) $GeoGAN_{noL_{shape}}$; (b) $GeoGAN_{noL_{cls}}$; (c) $GeoGAN_{noSamp}$. The two rows correspond to the images shown in the two rows of Figure~\ref{fig:segout1} (a).}
\label{fig:segout1_Abl}
\end{figure}

Table~\ref{tab:Abl} shows the segmentation results for different ablation studies. Figure~\ref{fig:segout1_Abl} shows the segmentation mask obtained by different baselines for the same image shown in Figure~\ref{fig:segout1} (a). The segmentation outputs are quite different from the ground truth and the one obtained by GeoGAN. In some cases the normal regions in the layers are included as pathological area, while parts of the infected region are not segmented with the pathological region. Either case is undesirable for disease diagnosis and quantification. Thus, different components of our cost functions are integral to the method's performance and excluding one or more of classification loss, geometric loss and sampling loss adversely affects segmentation performance.

\begin{table}[h]
 \begin{center}
\begin{tabular}{|c|c|c|c|}
\hline 
% & \multicolumn{4}{|c|}{Comparison approaches}  \\ \hline
{} & {GeoGAN}  & {GeoGAN}  & {GeoGAN} \\ 
{} & {$_{noL_{cls}}$}  & {$_{noL_{shape}}$}  & {$_{noSamp}$}\\ \hline
{DM} & {0.752(0.07)} & {0.759(0.09)} & {0.758(0.09)} \\ \hline
{HD} & {9.5(3.0)} & {9.2(3.3)}  & {9.0(3.2)}   \\ \hline
{MAE} & {0.078} & {0.080} & {0.079} \\  \hline
{p} & {0.001} & {0.001} & {0.0009} \\  \hline
% \hline
% {} & {GeoGAN}  & {GeoGAN}  & {GeoGAN} \\ 
% {} & {$_{onlyL_{cls}}$}  & {$_{onlyL_{shape}}$}  & {$_{onlySamp}$}\\ \hline
% %
% {DM} & {0.824(0.08)} & {0.825(0.07)} & {0.818(0.06)} \\ \hline
% {HD} & {11.2(2.9)} & {11.1(3.0)}  & {12.5(2.8)}   \\ \hline
\end{tabular}
\caption{Mean and standard deviation (in brackets) of segmentation results from ablation studies on COVID CT images from the $CTSeg1$ database. HD is in mm. $p-$values are with respect to GeoGAN.}
\label{tab:Abl}
\end{center}
\end{table}

\begin{figure}[h]
\centering
\begin{tabular}{ccc}
\includegraphics[height=2.5cm, width=2.5cm]{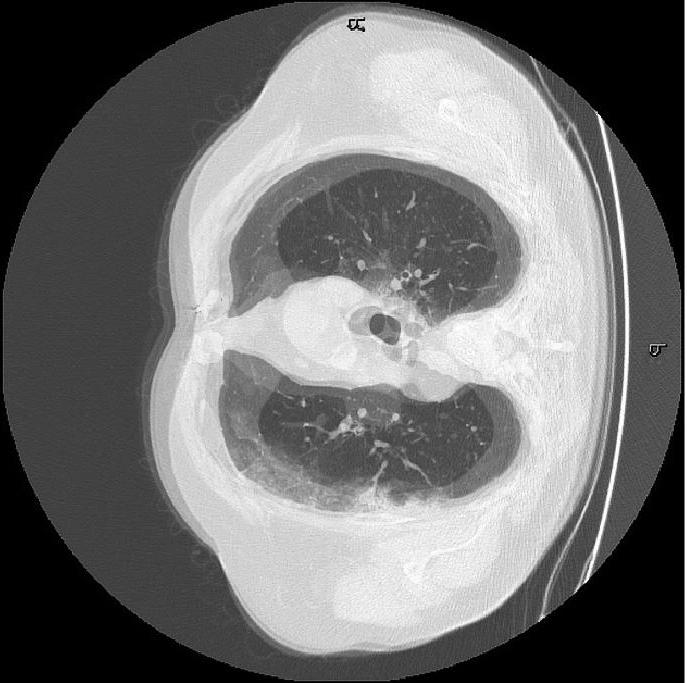} &
\includegraphics[height=2.5cm, width=2.5cm]{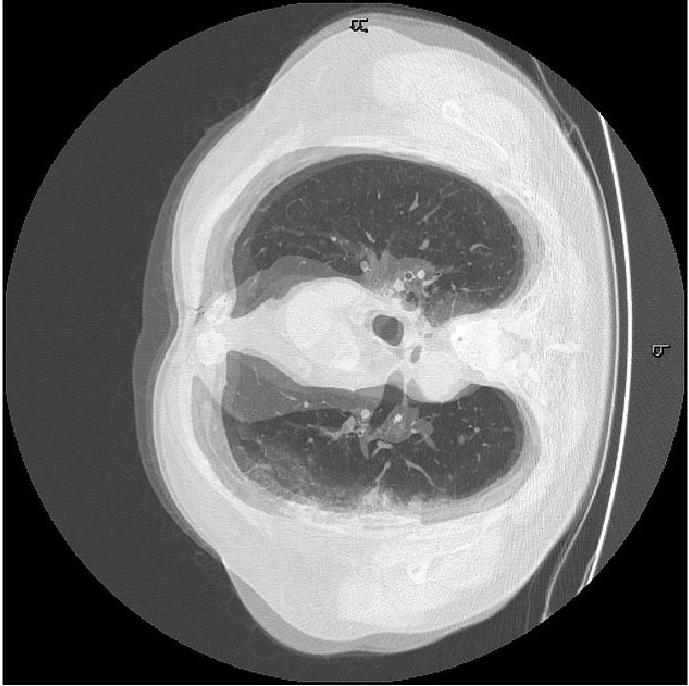} &
\includegraphics[height=2.5cm, width=2.5cm]{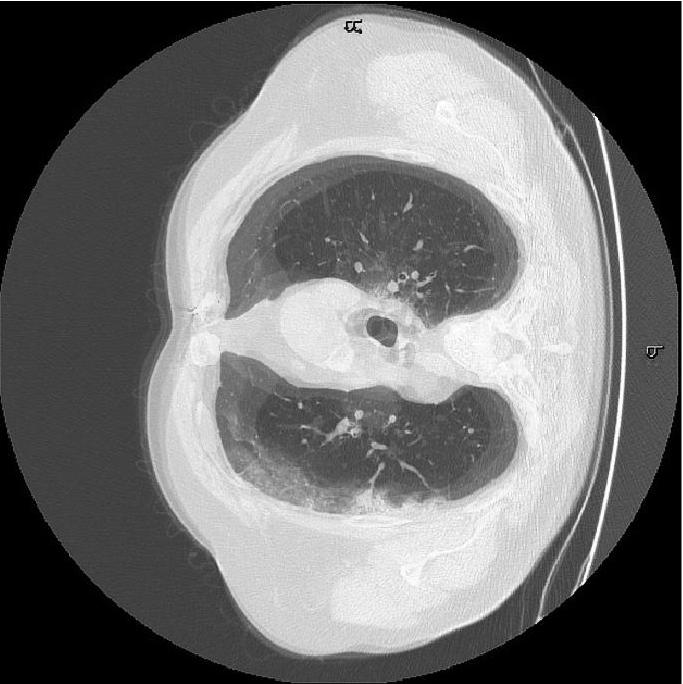} \\
(a) & (b) & (c) \\
\end{tabular}
\caption{ Generated images for ablation study methods: (a) $GeoGAN_{noL_{cls}}$; (b) $GeoGAN_{noL_{shape}}$; (c) $GeoGAN_{noSamp}$. The corresponding generated images for other methods are shown in Figure~\ref{fig:SynImages_comp}. }
\label{fig:SynImages_Abl}
\end{figure}

\subsection{Classification Results}

Table~\ref{tab:COVIDchal} summarizes the performance of different methods on the public challenge dataset of \cite{zhao2020COVID-CT-Dataset}\footnote{https://covid-ct.grand-challenge.org/Data/} consisting of $349$ CT images labeled as being COVID-19 positive. These
CT images have different sizes and come from $216$ patients. All of them are resized to $512\times512$. We train GeoGAN$_{WSS}$ on the training set, generate more images, train classifiers on the training images and apply it on the test set. 
The leaderboard can be accessed here\footnote{https://covid-ct.grand-challenge.org/Leaderboard/}. For all test submissions Accuracy (ACC), F1 score (F1) and area under curve (AUC) are calculated, while the ranking is based on the F1 score.
We use a DenseNet-121 architecture and employed GeoGAN$_{WSS}$ augmentation for the final results. Using conventional data augmentation we got the following values: F1=0.931,ACC=0.934,AUC=0.961, which would have placed us $6^{th}$ in the current leaderboard. However, by using GeoGAN$_{WSS}$ our results are ranked third although we obtain the highest AUC and ACC values, while being very close to the top ranked method in terms of F1 score. \textbf{For the completeness of the paper we would have liked to report the $p-$values at 95\% confidence. However we are unable to do so since we do not have access to the results of other methods.}

\begin{table}[h]
 \begin{center}
\begin{tabular}{|c|c|c|c|c|c|}
\hline 
{} & {Rank~1}  & {Rank~2} & {Rank~3}  & {Rank~4} & {Rank~5} \\ 
{} & {} & {} & {(GeoGAN$_{WSS}$)} & {} & {}\\ \hline
{ACC} & {0.939} & {0.953} & {\textbf{0.961}} & {0.956} & {0.945}\\ \hline
{F1} & {\textbf{0.967}} & {0.964} & {0.963} & {0.953} & {0.943} \\ \hline
{AUC} & {0.965} & {0.987} & {\textbf{0.991}} & {0.987} & {0.946} \\ \hline
\end{tabular}
\caption{Classification results on the public COVID-19 CT classification challenge dataset.}
\label{tab:COVIDchal}
\end{center}
\end{table}

In a second set of classification experiments we used the $CTSeg1$ dataset to generate augmentation images and train a classifier for detection COVID positive and negative cases. The classifier was used to classify images from the $CTSeg2$ and $CTSeg3$ datasets. Images from the two datasets were intensity  normalized and combined into one dataset. We add COVID negative images from the challenge dataset to get an almost equal distribution of positive and negative cases in the training and test sets.
 The classification results on the test set are summarized in Table~\ref{tab:ClassDenseNet} for DenseNet-121, and Table~\ref{tab:ClassResNet} for ResNet-50. 

An important component of our method is the WSS step which generates segmentation maps for a given image. For the classification dataset we do not have manual segmentation labels (from the challenge dataset) to verify the accuracy of our WSS method. However, the method's performance is reflected in the final classification performance which is better than other augmentation methods.

\begin{table*}[h]
 \begin{center}
\begin{tabular}{|c|c|c|c|c|c|c|c|}
\hline 
{} & {GeoGAN}  & {\cite{Zhao_CVPR2019}} & {DAGAN}  & {cGAN} & {GeoGAN$_{wClass}$} & {GeoGAN$_{wShape}$} & {GeoGAN$_{wSamp}$} \\  \hline
{Spe} & {\textbf{0.931}} & {0.916} & {0.893} & {0.881} & {0.873} & {0.869} & {0.877} \\ 
{} & {(0.023)} & {(0.028)} & {(0.031)} & {(0.024)} & {(0.030)} & {(0.035)} & {(0.038)} \\ \hline
{Sen} & {\textbf{0.942}} & {0.923} & {0.906} & {0.890} & {0.884} & {0.882} & {0.890} \\
{} & {(0.031)} & {(0.029)} & {(0.036)} & {(0.033)} & {(0.029)} & {(0.031)} & {(0.035)} \\ \hline
{Acc} & {\textbf{0.938}} & {0.920} & {0.902} & {0.887} & {0.881} & {0.877} & {0.886} \\ 
{} & {(0.026)} & {(0.030)} & {(0.033)} & {(0.036)} & {(0.032)} & {(0.034)} & {(0.034)} \\ \hline
{AUC} & {\textbf{0.967}} & {0.946} & {0.928} & {0.917} & {0.912} & {0.904} & {0.911} \\ 
{} & {(0.019)} & {(0.023)} & {(0.026)} & {(0.025)} & {(0.027)} & {(0.031)} & {(0.033)} \\ \hline
\end{tabular}
\caption{Classification results using the DenseNet-121 architecture. Values indicate mean(standard deviation). The best results are highlighted in bold.}
\label{tab:ClassDenseNet}
\end{center}
\end{table*}

\begin{table*}[h]
 \begin{center}
\begin{tabular}{|c|c|c|c|c|c|c|c|}
\hline 
{} & {GeoGAN}  & {\cite{Zhao_CVPR2019}} & {DAGAN}  & {cGAN} & {GeoGAN$_{wClass}$} & {GeoGAN$_{wShape}$} & {GeoGAN$_{wSamp}$} \\  \hline
{Spe} & {\textbf{0.915}} & {0.901} & {0.864} & {0.857} & {0.849} & {0.842} & {0.850} \\ 
{} & {(0.037)} & {(0.039)} & {(0.042)} & {(0.044)} & {(0.041)} & {(0.046)} & {(0.048)} \\ \hline
{Sen} & {\textbf{0.924}} & {0.907} & {0.872} & {0.864} & {0.855} & {0.851} & {0.858} \\
{} & {(0.036)} & {(0.040)} & {(0.041)} & {(0.045)} & {(0.043)} & {(0.047)} & {(0.046)} \\ \hline
{Acc} & {\textbf{0.921}} & {0.906} & {0.869} & {0.862} & {0.853} & {0.848} & {0.855} \\ 
{} & {(0.032)} & {(0.037)} & {(0.039)} & {(0.041)} & {(0.038)} & {(0.043)} & {(0.042)} \\ \hline
{AUC} & {\textbf{0.943}} & {0.927} & {0.887} & {0.883} & {0.876} & {0.870} & {0.877} \\ 
{} & {(0.024)} & {(0.028)} & {(0.029)} & {(0.031)} & {(0.035)} & {(0.034)} & {(0.035)} \\ \hline
\end{tabular}
\caption{Classification results using the ResNet-50 architecture. Values indicate mean(standard deviation). The best results are highlighted in bold.}
\label{tab:ClassResNet}
\end{center}
\end{table*}

\section{Conclusion}\label{sec:concl}

We propose a novel approach to generate plausible COVID-19 CT images by incorporating relationship between segmentation labels to guide the shape generation process.  Diversity is introduced in the image generation process through uncertainty sampling. Comparative results show that the augmented dataset from $GeoGAN$ outperforms standard data augmentation and other competing methods, when applied to segmentation of COVID-19 affected pathological regions in CT images. We show that synergy between shape, classification and sampling terms lead to improved segmentation and each of these terms is equally important in generating realistic shapes. % e relative contributions of each term and conclude that the shape prior term makes a significant contribution to the output, while a .
%  Our approach can be used for other medical imaging modalities without major changes to the workflow.
 
 Despite the good performance of our method we observe  failure cases when the base images are noisy due to inherent characteristics of the image acquisition procedure. %In future work we aim to evaluate our method's robustness on a wide range of medical imaging modalities such as MRI, Xray, etc. 
Our method is also useful to generate realistic images for educating  clinicians, where targeted synthetic images (e.g. generation of complex cases, or disease mimickers) can be used to speed-up training. %Similarly, the proposed approach could be used in quality control of deep learning systems to identify potential weaknesses through targeted high-throughput synthetic image generation and testing.

% {\small
% \bibliographystyle{ieee}
% \bibliography{CVPR2020_GeoGAN}
% }

% \bibliographystyle{model2-names.bst}\biboptions{authoryear}
\bibliographystyle{ieee_fullname}
\bibliography{COVID_MedIA_class_Ref,MyCitations_Conf,MyCitations_Journ}

\end{document}